\documentclass{iopjournal}
\usepackage{float}

\begin{document}

\articletype{Paper} %	 e.g. Paper, Letter, Topical Review...

\title{From isolated singular wells to exact Bloch bands: the trigonometric Pöschl–Teller lattice}

\author{Ulysses Camara da Silva$^1$\orcid{0000-0002-0409-5284}}

\affil{$^1$Department of Physics, Universidade Federal do Espírito Santo, Vit\'oria, Brazil}

%\affil{$^2$Department, Institution, City, Country}

%\affil{$^*$Author to whom any correspondence should be addressed.}

\email{ulysses.silva@ufes.br}


\begin{abstract}
\justifying
An exactly solvable lattice with periodic inverse-square singularities exhibits two physically distinct mechanisms of spectral-gap closure. The model arises from the generalized trigonometric Pöschl–Teller potential in the coupling range where both local solutions at each singularity are square-integrable. A transmitting, current-preserving self-adjoint matching condition connects neighboring cells and generates a nontrivial Bloch spectrum. For a generic lattice with two inequivalent singular sublattices, the spectrum contains two alternating families of gaps, and the band edges reproduce the scale-invariant spectra of the associated finite singular system. One gap family closes when the two sublattices become equivalent. In this symmetric limit, however, the primitive period halves, so the apparent band crossing results from Brillouin-zone folding. A different family closes at a critical but genuinely asymmetric configuration. Because the primitive period is preserved, adjacent bands then meet at the center of the original Brillouin zone in a genuine massless Dirac-like cone. Near either critical limit, a small finite gap acts as an effective Dirac mass: the dispersion remains quadratic near a band edge but crosses over to a massive Dirac-like form beyond that narrow region. The band curvatures define effective electron and hole masses. Exact Bloch eigenfunctions further reveal that isospectral lattices can display different single-particle probability densities. This model therefore provides an analytically controlled one-dimensional setting in which conventional quadratic carriers, massive Dirac-like band dispersions, and a tunable massless contact coexist within the same singular lattice.
\end{abstract}

\section{Introduction}
\label{sec:introduction}

Exactly solvable models occupy a special place in quantum mechanics. Besides providing benchmarks for approximation schemes and numerical calculations, they often expose physical mechanisms that are difficult to isolate in more realistic systems. Nevertheless, the number of one-dimensional potentials for which the stationary Schr\"odinger equation can be solved analytically is limited, and the situation is even more restrictive for periodic systems. Only a few continuous one-dimensional lattices admit simultaneous analytical control of their Bloch spectrum, eigenfunctions, and dispersion relations.
Models such as the Kronig--Penney and Lam\'e systems illustrate how exceptional this degree of solvability is~\cite{KronigPenney1931,GesztesyWeikard1995}. Exact periodic models are therefore valuable for studying the formation and closing of spectral gaps, the effective dynamics of charge carriers, and the spatial structure of Bloch states.

Among the known solvable quantum systems, an important subclass contains inverse-square singularities. Such interactions arise in conformal quantum mechanics, Calogero-type systems, radial reductions of higher-dimensional problems, black-hole perturbations, and fluctuations in asymptotically anti-de Sitter geometries~\cite{deAlfaro:1976vlx,Calogero:1970nt,Ferrari:1984zz,Ishibashi:2004wx}. They also present a spectral subtlety that is absent from regular potentials. In the neighborhood of a singular point \(x=x_i\), one has
\begin{equation}
 \frac{2m}{\hbar^2}V(x)\simeq \frac{g_i}{(x-x_i)^2},
 \qquad
 \psi(x)\simeq
 A_i\lvert x-x_i\rvert^{\frac12+\nu_i}
 +B_i\lvert x-x_i\rvert^{\frac12-\nu_i},
 \qquad
 \nu_i=\sqrt{\frac14+g_i}.
 \label{eq:intro_local_behavior}
\end{equation}
For \(0<\nu_i<1\), both independent solutions are locally square-integrable.
Normalizability then fails to select a unique boundary condition and, by
itself, does not define the Hamiltonian or its spectrum~\cite{Efimov_Grif,Gitman,renor_m_w_c}. An additional physical
prescription is required to relate the asymptotic coefficients on the two sides
of the singularity. In operator language, this amounts to choosing a
self-adjoint realization of the Hamiltonian; locally, the same requirement may
be expressed as conservation of the probability current, or of the associated
Wronskian form.

\begin{figure}[t]
 \centering
 \subfigure[]{
  \includegraphics[scale=0.55]{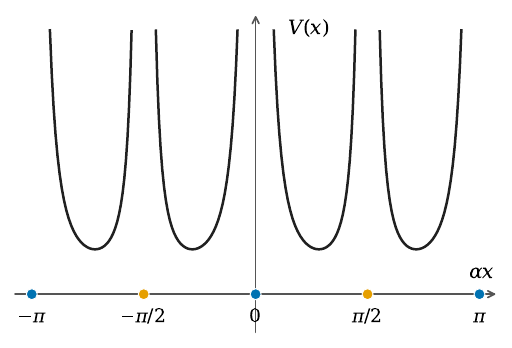}
  \label{fig:potential_repulsive}
 }
 \subfigure[]{
  \includegraphics[scale=0.55]{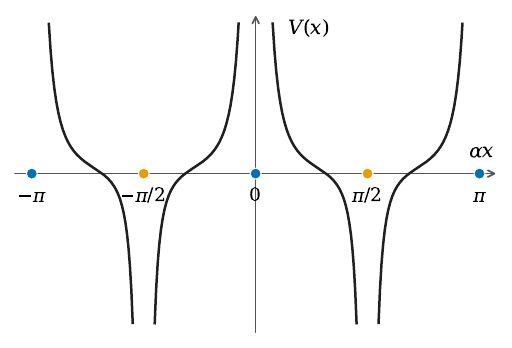}
  \label{fig:potential_mixed}
 }
 \subfigure[]{
  \includegraphics[scale=0.55]{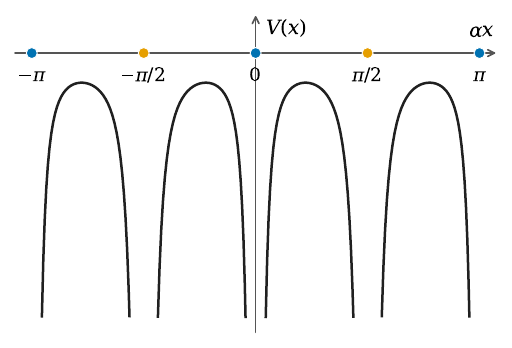}
  \label{fig:potential_attractive}
 }
 \caption{Representative forms of the periodic generalized trigonometric
 P\"oschl--Teller potential over two primitive periods: (a) two repulsive
 families of singularities, \(g_s,g_c>0\); (b) alternating repulsive and
 attractive singularities, \(g_sg_c<0\); and (c) two attractive families of
 singularities, \(g_s,g_c<0\). Blue and orange markers identify the singular
 sites associated with \(g_s\) and \(g_c\), respectively.}
 \label{fig:potential_classes}
\end{figure}

In this work, we investigate the generalized trigonometric P\"oschl--Teller
potential~\cite{Poschl:1933zz}
\begin{equation}
 \frac{2m}{\hbar^2}V(x)
 =\alpha^2\left[
 \frac{g_s}{\sin^2(\alpha x)}
 +\frac{g_c}{\cos^2(\alpha x)}
 \right],
 \qquad
 g_s=\nu_s^2-\frac14,
 \qquad
 g_c=\nu_c^2-\frac14,
 \label{eq:intro_TPT_potential}
\end{equation}
where \(\alpha^{-1}\) sets the characteristic length scale. Depending on the
signs of the two couplings, the periodic potential contains two repulsive
families of singularities, alternating attractive and repulsive singularities,
or two attractive families. Representative profiles of these three cases are
shown in figures~\ref{fig:potential_classes}(a),
\ref{fig:potential_classes}(b), and \ref{fig:potential_classes}(c),
respectively. On the elementary interval
\(0<x<\pi/(2\alpha)\), the potential is bounded by two inverse-square
singularities. Its finite-interval spectrum, the admissible boundary
conditions in the different coupling regions, and the construction of a
singular double well were analyzed in our previous work
\cite{CamaraDaSilvaTPT}.

The periodic interpretation follows directly from the identities~\cite{NIST:DLMF}
\begin{equation}
 \frac{\alpha^2}{\sin^2(\alpha x)}
 =\sum_{n\in\mathbb Z}
 \frac{1}{\left(x-\frac{n\pi}{\alpha}\right)^2},
 \qquad
 \frac{\alpha^2}{\cos^2(\alpha x)}
 =\sum_{n\in\mathbb Z}
 \frac{1}{\left[x-\frac{(n+\frac12)\pi}{\alpha}\right]^2}.
 \label{eq:intro_lattice_series}
\end{equation}
The two trigonometric terms thus generate interlaced arrays of inverse-square
centers: the \(g_s\) sites lie at \(x=n\pi/\alpha\), while the \(g_c\) sites
lie at \(x=(n+1/2)\pi/\alpha\). For \(g_s\neq g_c\), the two sublattices are
inequivalent and the primitive period is \(L=\pi/\alpha\), so the model may be
viewed as a one-dimensional diatomic lattice. When \(g_s=g_c\), the two
sublattices become equivalent, the primitive period is reduced to
\(L/2=\pi/(2\alpha)\), and the lattice is effectively monatomic.

This periodic continuation requires some care. In the weak-to-medium coupling
region
\begin{equation}
 0<\nu_s,\nu_c<1,
 \qquad\text{equivalently}\qquad
 -\frac14<g_s,g_c<\frac34,
 \label{eq:intro_coupling_region}
\end{equation}
supersymmetric quantum mechanics provides one possible prescription for
resolving the ambiguity in equation~\eqref{eq:intro_local_behavior}. The usual
SUSY-compatible domain selects the less singular local solution and eliminates
the second asymptotic coefficient
\cite{Asim_paper,Lima:2019xzg,Gango_book}. In the
language of self-adjoint extensions and renormalization, this prescription
selects a scale-invariant ultraviolet fixed point; generic extensions need not
be compatible with the same supersymmetric realization
\cite{Falomir:2005yw}. If this prescription is imposed independently
at every pole of the periodic potential, each singularity is an impenetrable
barrier. The cells remain dynamically disconnected and the periodic repetition
merely reproduces the discrete spectrum of an isolated interval, rather than a
genuine dispersive band structure.

Here we choose a different self-adjoint realization. Both square-integrable
asymptotic behaviors are retained, and their coefficients are matched across
each singularity so that the boundary form - and hence the probability
current - is continuous. This transmitting condition couples neighboring
intervals and allows quantum propagation over the entire real line. Although
the differential expression remains locally shape invariant inside every open
interval, the transmitting domain is not the conventional one selected by
SUSY, and operator-level supersymmetry should not be assumed. The construction
extends the singular double well of Ref.~\cite{CamaraDaSilvaTPT}: the finite
two-well system becomes the elementary building block of an exactly solvable
one-dimensional lattice.

The matching conditions determine a unimodular transfer matrix over one primitive period. Combining its eigenvalue equation with Bloch's condition yields the exact dispersion relation. The band edges form two alternating families and coincide with the two scale-invariant fixed-point spectra of the finite-interval problem. This connection explains how the discrete levels of the singular double well become the endpoints of the Bloch bands when the elementary system is repeated over the real line. It also reveals two independent mechanisms for closing spectral gaps.

The first mechanism occurs as \(\lvert\nu_s-\nu_c\rvert\to0\). The gap
controlled by the sublattice asymmetry becomes small and the corresponding
avoided crossing acquires a massive Dirac-like form. At exact equality, however,
the primitive cell is halved and the Brillouin zone doubles. The apparent
crossing obtained by retaining the larger, asymmetric cell is therefore a
folded-band crossing, not a genuine contact between distinct bands. The second mechanism occurs on the complementary-coupling line $\nu_s+\nu_c=1$, $\nu_s\neq\nu_c$. In this case the original primitive period remains unchanged, a different family of gaps closes at the center of the Brillouin zone, and adjacent bands meet in a genuine massless Dirac-like cone. Thus the two conditions \(\nu_s=\nu_c\) and \(\nu_s+\nu_c=1\) have sharply different physical meanings, even though each removes one family of gaps while leaving the other one finite.

Although the underlying Schr\"odinger problem remains nonrelativistic, the local two-branch dispersion near a closing gap has the same functional form as a one-dimensional massive Dirac spectrum, up to a common quadratic correction. At the critical point the gap vanishes, and the two branches meet linearly. Throughout this work, the terms ``massive'' and ``massless Dirac-like'' refer to this emergent band dispersion, rather than to a fundamental relativistic Hamiltonian or to an intrinsic spinor degree of freedom. Similar effective dispersions can be engineered in one-dimensional quantum-simulation platforms
~\cite{GarreauZehnle2017,ZeunerEtAl2012}.
%Away from these gap-closing conditions, the band extrema have the conventional quadratic dispersion and define effective electron and hole masses. Near a small but nonzero gap, the exact spectrum displays a crossover from this nonrelativistic behavior in the immediate vicinity of the edge to a massive Dirac-like regime at larger momentum deviations. The gap supplies the effective Dirac mass, while a common quadratic term describes the leading departure from an ideal relativistic spectrum. On the line \(\nu_s+\nu_c=1\), the mass term vanishes and the linear contact survives in the original Brillouin zone. In this restricted but precise sense, the model provides an analytically controllable one-dimensional analogue of tunable massive and massless Dirac quasiparticles~\cite{GarreauZehnle2017,ZeunerEtAl2012}.

Exact knowledge of the transfer-matrix eigenvectors also fixes the relative coefficient of the two independent solutions in each cell. After normalization over one primitive period, the full Bloch probability density can therefore be obtained analytically. The density varies smoothly, and generally without a qualitative rearrangement, as the quasimomentum moves within a given band; the more pronounced redistribution occurs when a gap is crossed and the next band is entered. In an asymmetric cell, this alternation reflects the different local behaviors allowed at the two singular sublattices and the preferential accumulation near the more attractive, or less repulsive, singularity. Moreover, the discrete transformation
\begin{equation}
 (\nu_s,\nu_c)\longmapsto(1-\nu_s,1-\nu_c)
 \label{eq:intro_isospectral_map}
\end{equation}
leaves the energy bands invariant but generally changes the Bloch eigenfunctions and their probability densities. The band spectrum alone therefore does not fully characterize the spatial physics of the lattice.

The paper is organized as follows. In Sec.~II, we solve the stationary Schr\"odinger equation in an elementary interval and derive the asymptotic bases at its singular endpoints. In Sec.~III, the current-preserving matching conditions are used to construct the transfer matrix and obtain the exact
Bloch equations for the asymmetric and symmetric lattices. Section~IV examines the band edges, widths, gaps, and discrete spectral symmetries, and relates the edge energies to the fixed-point spectra of the finite problem. In Sec.~V, we derive the band-edge expansions and effective electron and hole masses, analyze the massive Dirac crossover near a small gap, and distinguish the folded crossing of the symmetric lattice from the genuine massless contact at
\(\nu_s+\nu_c=1\). Section~VI determines the Bloch eigenvectors and studies the normalized single-particle probability densities for generic, symmetric, and complementary-coupling lattices. Finally, Sec.~VII summarizes the results and discusses the use of the same inverse-square matching mechanism in other exactly solvable singular lattices and in effective one-dimensional quantum
simulation settings.

\section{Exact solution in an elementary interval}
\label{sec:cell_solution}

Having introduced the periodic potential, we now solve the stationary Schrödinger equation in the elementary interval
$\mathcal{I}_{0}=\left(0,\frac{\pi}{2\alpha}\right)$, bounded by a \(g_s\) singularity at \(x=0\) and a \(g_c\) singularity at $x=\pi/(2\alpha)$. In the asymmetric case, this interval constitutes one half of the primitive cell, whose length is \(\pi/\alpha\). The continuation to the adjacent interval and the matching across the singular points are discussed in section~\ref{sec:matching}.

The stationary Schr\"odinger equation becomes
\begin{equation}
\left[
\frac{d^2}{dx^2}
+k^2-\alpha^2\left(\frac{\nu_s^2-\frac14}{\sin^2(\alpha x)}
+\frac{\nu_c^2-\frac14}{\cos^2(\alpha x)}\right)
\right]\psi_k(x)
=0,
\qquad
k^2=\frac{2mE}{\hbar^2}.
\label{eq:schrodinger_cell}
\end{equation}
Throughout this work, the parameters are restricted to
\[
0<\nu_s,\nu_c<1,
\]
so that both independent local solutions are square-integrable at each singular endpoint. A convenient form of the general solution to equation~(\ref{eq:schrodinger_cell}) in the elementary interval is \cite{CamaraDaSilvaTPT}
\begin{eqnarray}
&&\psi_k(x)=
\frac{A_0(k)}
{\sqrt{2\nu_s}\,\alpha^{\frac12+\nu_s}}
\bigl[\tan(\alpha x)\bigr]^{\frac12+\nu_s}
\bigl[\cos(\alpha x)\bigr]^{-k/\alpha}
\mathcal F\left(\nu_s,\nu_c;k;-\tan^2(\alpha x)\right)
\nonumber\\
&&+\frac{B_0(k)}
{\sqrt{2\nu_s}\,\alpha^{\frac12-\nu_s}}
\bigl[\tan(\alpha x)\bigr]^{\frac12-\nu_s}
\bigl[\cos(\alpha x)\bigr]^{-k/\alpha}
\mathcal F\left(-\nu_s,\nu_c;k;-\tan^2(\alpha x)\right),
\label{eq:general_solution_first_interval}
\end{eqnarray}
where
\begin{eqnarray}
\mathcal F(\nu_s,\nu_c;k;z)
\equiv
{}_2F_1\left(
\frac12+\frac{\nu_s}{2}-\frac{\nu_c}{2}+\frac{k}{2\alpha},
\frac12+\frac{\nu_s}{2}+\frac{\nu_c}{2}+\frac{k}{2\alpha};
1+\nu_s;
z
\right).
\label{eq:hypergeometric_definition}
\end{eqnarray}
The normalization factors in equation~(\ref{eq:general_solution_first_interval}) have been chosen so that, as $x\rightarrow0^+$,
\begin{equation}
\psi_k(x)
\simeq
A_0(k)\frac{x^{\frac12+\nu_s}}{\sqrt{2\nu_s}}
+
B_0(k)\frac{x^{\frac12-\nu_s}}{\sqrt{2\nu_s}}.
\label{eq:asymptotic_left}
\end{equation}
Defining the local basis
\begin{equation}
v_a^{(0)}(x)=
\frac{x^{\frac12+\nu_s}}{\sqrt{2\nu_s}},
\qquad
v_b^{(0)}(x)=
\frac{x^{\frac12-\nu_s}}{\sqrt{2\nu_s}},
\label{eq:basis_left}
\end{equation}
we obtain the normalized Wronskian
\begin{equation}
W[v_b^{(0)},v_a^{(0)}]
=v_b^{(0)}v_a^{(0)\prime}-v_b^{(0)\prime}v_a^{(0)}=1.
\label{eq:wronskian_left}
\end{equation}

At the opposite endpoint, $x\rightarrow\frac{\pi}{2\alpha}^-$, the wave function behaves as
\begin{equation}
\psi_k(x)
\simeq
A_1(k)\frac{1}{\sqrt{2\nu_c}}
\left(\frac{\pi}{2\alpha}-x\right)^{\frac12+\nu_c}
-
B_1(k)\left[-\frac{1}{\sqrt{2\nu_c}}
\left(\frac{\pi}{2\alpha}-x\right)^{\frac12-\nu_c}\right],
\label{eq:asymptotic_right}
\end{equation}
where the coefficients at the two endpoints are related by
\begin{eqnarray}
\frac{A_{1}(k)}{\alpha^{\nu_c}}=&&\sqrt{\frac{\nu_c}{\nu_s}}\Bigg(
\frac{A_{0}(k)}{\alpha^{\nu_s}}
\frac{\Gamma(-\nu_c)\Gamma(1+\nu_s)}
{\Gamma\left(\frac{1}{2}+\frac{\nu_s}{2}-\frac{\nu_c}{2}+\frac{k}{2\alpha}\right)
 \Gamma\left(\frac{1}{2}+\frac{\nu_s}{2}-\frac{\nu_c}{2}-\frac{k}{2\alpha}\right)}
\nonumber\\
&&+\frac{B_{0}(k)}{\alpha^{-\nu_s}}
\frac{\Gamma(-\nu_c)\Gamma(1-\nu_s)}
{\Gamma\left(\frac{1}{2}-\frac{\nu_s}{2}-\frac{\nu_c}{2}+\frac{k}{2\alpha}\right)
 \Gamma\left(\frac{1}{2}-\frac{\nu_s}{2}-\frac{\nu_c}{2}-\frac{k}{2\alpha}\right)}
\Bigg),
\nonumber\\
\frac{B_{1}(k)}{\alpha^{-\nu_c}}=&&\sqrt{\frac{\nu_c}{\nu_s}}\Bigg(
\frac{A_{0}(k)}{\alpha^{\nu_s}}
\frac{\Gamma(\nu_c)\Gamma(1+\nu_s)}
{\Gamma\left(\frac{1}{2}+\frac{\nu_s}{2}+\frac{\nu_c}{2}+\frac{k}{2\alpha}\right)
 \Gamma\left(\frac{1}{2}+\frac{\nu_s}{2}+\frac{\nu_c}{2}-\frac{k}{2\alpha}\right)}
\nonumber\\
&&+\frac{B_{0}(k)}{\alpha^{-\nu_s}}
\frac{\Gamma(\nu_c)\Gamma(1-\nu_s)}
{\Gamma\left(\frac{1}{2}-\frac{\nu_s}{2}+\frac{\nu_c}{2}+\frac{k}{2\alpha}\right)
 \Gamma\left(\frac{1}{2}-\frac{\nu_s}{2}+\frac{\nu_c}{2}-\frac{k}{2\alpha}\right)}
\Bigg).
\label{master}
\end{eqnarray}
The minus sign multiplying $B_1$ in equation~(\ref{eq:asymptotic_right}) is incorporated into the second element of the local basis,
\begin{equation}
v_a^{(1)}(x)=
\frac{1}{\sqrt{2\nu_c}}
\left(\frac{\pi}{2\alpha}-x\right)^{\frac12+\nu_c},
\qquad
v_b^{(1)}(x)=
-\frac{1}{\sqrt{2\nu_c}}
\left(\frac{\pi}{2\alpha}-x\right)^{\frac12-\nu_c},
\label{eq:basis_right}
\end{equation}
so that the Wronskian retains the same normalization, $W[v_b^{(1)},v_a^{(1)}]=1$. At this stage, the choice of local bases with the same unit Wronskian may appear merely conventional. It will, however, make the matching across adjacent intervals particularly transparent. Its physical role becomes clear in section~\ref{sec:matching}, where the matching rule is derived from the boundary form associated with self-adjointness of the Hamiltonian.%and probability conservation.

Introducing the dimensionless energy parameter $\kappa=k/\alpha$, the relations in equation~(\ref{master}) can be written as
\begin{equation}
\left(
\begin{array}{c}
\alpha^{-\nu_c}A_1(k)\\
\alpha^{\nu_c}B_1(k)
\end{array}
\right)
=
\mathcal{C}_{sc}(\kappa)
\left(
\begin{array}{c}
\alpha^{-\nu_s}A_0(k)\\
\alpha^{\nu_s}B_0(k)
\end{array}
\right),
\label{eq:connection_matrix_definition}
\end{equation}
where
\begin{equation}
 \mathcal C_{sc}(\kappa)=
 \begin{pmatrix}
  C^{sc}_{11}(\kappa)&C^{sc}_{12}(\kappa)\\
  C^{sc}_{21}(\kappa)&C^{sc}_{22}(\kappa)
 \end{pmatrix}.
 \label{Matriz_C}
\end{equation}
Thus, $\mathcal C_{sc}$ contains all the information required to express the
two asymptotic behaviors at the $g_s$ endpoint in terms of the corresponding
behaviors at the $g_c$ endpoint. Its four entries, written directly in terms
of $\nu_s$ and $\nu_c$, are collected in
appendix~\ref{app:transfer_eigenvectors}. A direct calculation using the
elementary Gamma-function identities
\begin{equation}
 \Gamma(1+z)=z\Gamma(z),
 \qquad
 \Gamma(z)\Gamma(1-z)=\frac{\pi}{\sin(\pi z)}
 \label{Gama}
\end{equation}
shows that, with the oriented local bases used above,
\begin{equation}
 \det\mathcal C_{sc}(\kappa)=-1.
 \label{eq:det_connection_matrix}
\end{equation}
The symmetric lattice corresponds to the limit $\nu_s=\nu_c\equiv\nu$, in
which $\mathcal C_{sc}$ reduces to the matrix
$\mathcal C_{\mathrm{sym}}(\nu;\kappa)$ given explicitly in
appendix~\ref{app:transfer_eigenvectors}. The simplification of its diagonal
elements follows directly from the Gamma-function identities in
equation~(\ref{Gama}).

%---------------------------

\section{Matching conditions and Bloch bands}
\label{sec:matching}

We now extend the solution obtained in the preceding section across the singular point at $x=\pi/(2\alpha)$. In the second half of the primitive cell, $\pi/(2\alpha)<x<\pi/\alpha$, the potential is obtained from that in the elementary interval by a translation of $\pi/(2\alpha)$, under which
$$\sin(\alpha x)\rightarrow \cos(\alpha x), \ \ \cos(\alpha x)\rightarrow-\sin(\alpha x).$$
Consequently, the two couplings are interchanged. In terms of a new pair of independent coefficients, the general solution in this interval can be written as
\begin{eqnarray}
     \psi_k(x)={}&
 \frac{\widetilde A_1(k)}{\sqrt{2\nu_c}\,
 \alpha^{1/2+\nu_c}}
 [-\cot(\alpha x)]^{1/2+\nu_c}[\sin(\alpha x)]^{-\frac{k}{\alpha}}
 \mathcal F\!\left(\nu_c,\nu_s;k;-\cot^2(\alpha x)\right)
 \nonumber\\
 &+\frac{\widetilde B_1(k)}{\sqrt{2\nu_c}\,
 \alpha^{1/2-\nu_c}}
 [-\cot(\alpha x)]^{1/2-\nu_c}[\sin(\alpha x)]^{-\frac{k}{\alpha}}
 \mathcal F\!\left(-\nu_c,\nu_s;k;-\cot^2(\alpha x)\right).
 \label{eq:solution_second_half}
\end{eqnarray}

As $x\rightarrow\frac{\pi}{2\alpha}^+$,
equation (\ref{eq:solution_second_half}) gives
\begin{equation}
 \psi_k(x)\simeq
 \widetilde A_1(k)\frac{1}{\sqrt{2\nu_c}}\left(x-\frac{\pi}{2\alpha}\right)^{1/2+\nu_c}
 +\widetilde B_1(k)\frac{1}{\sqrt{2\nu_c}}\left(x-\frac{\pi}{2\alpha}\right)^{1/2-\nu_c}.
 \label{eq:asymptotic_middle_right}
\end{equation}
The corresponding local basis,
\begin{equation}
 \widetilde v_a^{(1)}(x)=\frac{1}{\sqrt{2\nu_c}}\left(x-\frac{\pi}{2\alpha}\right)^{1/2+\nu_c},
 \qquad
 \widetilde v_b^{(1)}(x)=\frac{1}{\sqrt{2\nu_c}}\left(x-\frac{\pi}{2\alpha}\right)^{1/2-\nu_c},
 \label{eq:middle_right_basis}
\end{equation}
is again normalized according to $W[\widetilde v_b^{(1)},\widetilde v_a^{(1)}]=1$. To establish the matching condition, consider two functions $\psi_k$ and $\psi_{k'}$ in the domain of the lattice Hamiltonian. The differential expression is defined separately on the open intervals between successive singularities, namely
\begin{eqnarray}
    \bigcup_{n\in\mathbb{Z}}\left(\frac{n\pi}{2\alpha},\frac{(n+1)\pi}{2\alpha}\right).
\end{eqnarray}
Self-adjointness requires the boundary contributions at every internal singularity to cancel. For example, on the punctured cell
$(0,\pi/(2\alpha))\cup(\pi/(2\alpha),\pi/\alpha)$, the boundary form must reduce to contributions from the two outer endpoints only:
\begin{equation}
 \mathcal{B}[\psi_{k'},\psi_k]
 = -\frac{2m}{\hbar^2}
 \left[\big<\psi_{k'}|H\psi_k\big>-
 \langle H\psi_{k'}|\psi_k\rangle\right]= J_{k',k}\left(\frac{\pi}{\alpha}^-\right)-J_{k',k}\left(0^+\right),
 \label{eq:self_adjoint_boundary_form}
\end{equation}
where
\begin{equation}
 J_{k',k}(x)=\psi_{k'}^*(x)\psi_k'(x)-\psi_{k'}^{*\prime}(x)\psi_k(x).\label{eq:sesquilinear_current}
\end{equation}
Thus, within a single cell, $\mathcal{B}[\psi_{k'},\psi_k]$ may contain only terms associated with the outer boundaries; all contributions from internal singularities must cancel. When neighboring cells are added, their common endpoints become internal boundaries and cancel in the same way. Since $x=\pi/(2\alpha)$ is excluded from the differential domain, integration by parts must be performed separately on the two open intervals. Therefore,%Introducing a cutoff $\varepsilon>0$, one obtains
\begin{eqnarray}
 \mathcal{B}[\psi_{k'},\psi_k]
 &=&
 \left[
 \int_{0^+}^{\frac{\pi}{2\alpha}^-}
 +
 \int_{\frac{\pi}{2\alpha}^+}^{\frac{\pi}{\alpha}^-}
 \right]
 \frac{dJ_{k',k}(x)}{dx}\,dx,
 \nonumber\\
 &=&\bigg[
 J_{k',k}\left(\frac{\pi}{2\alpha}^-\right)
 -
 J_{k',k}\left(\frac{\pi}{2\alpha}^+\right)
 \bigg]+ J_{k',k}\left(\frac{\pi}{\alpha}^-\right)-J_{k',k}\left(0^+\right).
 \label{eq:punctured_cell_boundary_form}
\end{eqnarray}
The two terms inside square brackets must cancel. Substitution of equations~(\ref{eq:asymptotic_right}) and~(\ref{eq:asymptotic_middle_right}) then gives
\begin{align}
 A_1^*(k')B_1(k)-B_1^*(k')A_1(k)=\widetilde B_1^*(k')\widetilde A_1(k)
   -\widetilde A_1^*(k')\widetilde B_1(k).
\label{eq:currents_at_central_singularity}
\end{align}
The most general matching condition may mix the two asymptotic coefficients. Here we select the simplest prescription: it is energy independent, does not mix the two local behaviors, and is represented by a real matrix. The equality of the boundary currents is then satisfied by
\begin{eqnarray}
 \widetilde A_1(k)=-A_1(k),\qquad
 \widetilde B_1(k)=B_1(k).
 \label{eq:transparent_matching}  
\end{eqnarray}
This choice satisfies equation~(\ref{eq:currents_at_central_singularity})
and defines the transmitting self-adjoint realization considered below. It
also has a direct physical interpretation inherited from the isolated
singular double-well problem studied in Ref.~\cite{CamaraDaSilvaTPT}. The
more regular behavior, $|x-x_i|^{1/2+\nu_i}$, always vanishes at the
singularity and is naturally continued antisymmetrically, whereas the less
regular behavior, $|x-x_i|^{1/2-\nu_i}$, is continued symmetrically. The
relations $\widetilde A_i=-A_i$ and $\widetilde B_i=B_i$ therefore identify
the $A$ and $B$ sectors with the locally odd and even extensions,
respectively. This interpretation becomes particularly transparent in the
free limit $\nu_i\rightarrow1/2$, where the singular coupling
$g_i=\nu_i^2-1/4$ vanishes. The two local behaviors then reduce to $|x-x_i|$
and a constant; after matching, they become the usual odd Dirichlet and even
Neumann solutions. Thus the prescription adopted here is the periodic
continuation of the parity-preserving regularization used in the finite
double-well problem, rather than an algebraic choice made specifically for
the lattice.

The matching is represented by
$\eta=\operatorname{diag}(-1,1)$ and should not be confused with prescribing
an energy-independent ratio $B_i/A_i$ at an isolated singularity. Here the
energy-independent short-distance input is the relation between the
coefficients on opposite sides of each singular point. The ratio $B_i/A_i$
of a Bloch state is instead determined globally by propagation through the
elementary cell together with Bloch's condition, and consequently depends
on the energy and the Bloch momentum.
%which obeys equation~(\ref{eq:currents_at_central_singularity}) and preserves both time-reversal and parity symmetry. This matching is represented by the matrix $\eta=\operatorname{diag}(-1,1)$.

The asymptotic analysis as $x\to\pi/\alpha^-$ is identical to that of
section~\ref{sec:cell_solution}, with $\nu_s\leftrightarrow\nu_c$. Hence the
connection matrix for the second half of the cell is
\begin{equation}
 \begin{pmatrix}
  \alpha^{-\nu_s}\widetilde A_2(k)\\
  \alpha^{\nu_s}\widetilde B_2(k)
 \end{pmatrix}
 =\mathcal C_{cs}(\kappa)
 \begin{pmatrix}
  \alpha^{-\nu_c}\widetilde A_1(k)\\
  \alpha^{\nu_c}\widetilde B_1(k)
 \end{pmatrix}.
 \label{eq:connection_matrix_definition_2}
\end{equation}
Here $\mathcal C_{cs}$ is obtained from $\mathcal C_{sc}$ by the interchange
$\nu_s\leftrightarrow\nu_c$; its explicit elements are also given in
appendix~\ref{app:transfer_eigenvectors}.

At $x=\pi/\alpha$, the same matching rule connects the second half-cell to the
next primitive cell. Therefore, the transfer matrix over one asymmetric
primitive period, $L=\pi/\alpha$, is
\begin{equation}
 \mathcal T(\kappa)
 =\eta\mathcal C_{cs}(\kappa)\eta\mathcal C_{sc}(\kappa)
 =\begin{pmatrix}
   \mathcal T_{11}&\mathcal T_{12}\\
   \mathcal T_{21}&\mathcal T_{22}
  \end{pmatrix}.
 \label{eq:transfer_matrix}
\end{equation}
Because $\det\eta=\det\mathcal C_{sc}=\det\mathcal C_{cs}=-1$, it follows
that $\det\mathcal T=1$. Moreover, direct multiplication gives
$\mathcal T_{11}=\mathcal T_{22}$. Explicit Gamma-function expressions for
all its entries are given in appendix~\ref{app:transfer_eigenvectors}.

Bloch's condition,

$$\psi_{k,Q}\left(x+L\right)=e^{iQL}\psi_{k,Q}(x),\qquad L=\frac{\pi}{\alpha},$$
applies equally to the vector of asymptotic coefficients and gives the eigenvalue equation
\begin{equation}
 \mathcal T\left(
\begin{array}{c}
\alpha^{-\nu_s}A_0(k)\\
\alpha^{\nu_s}B_0(k)
\end{array}
\right)=e^{iQL}\left(
\begin{array}{c}
\alpha^{-\nu_s}A_0(k)\\
\alpha^{\nu_s}B_0(k)
\end{array}
\right).
 \label{eq:bloch_eigenvalue}
\end{equation}
The characteristic equation becomes
\begin{equation}
 e^{2iQL}-\left(\operatorname{tr}\mathcal T\right)e^{iQL}+1=0,
 \label{eq:transfer_characteristic}
\end{equation}
and hence
\begin{equation}
 \cos(QL)=\Delta(\kappa),
 \qquad
 \Delta(\kappa)\equiv\frac{1}{2}\operatorname{tr}\mathcal T.
 \label{eq:discriminant}
\end{equation}
For the present model, multiplying the two connection matrices and using the Gamma-function identities in equation~(\ref{Gama}) yields
\begin{eqnarray}
 \Delta(\kappa)
 =\frac{\cos(\pi\kappa)+\cos(\pi\nu_s)\cos(\pi\nu_c)}
 {\sin(\pi\nu_s)\sin(\pi\nu_c)}\label{discriminante_a}
\end{eqnarray}
Equation~(\ref{discriminante_a}) is the exact band equation for the asymmetric lattice. Since $Q$ is real, the allowed bands are determined by the inequality
\begin{equation}
 |\Delta(\kappa)|\leq 1
 \qquad\Longleftrightarrow\qquad
 |\operatorname{tr}\mathcal T|\leq2,
 \label{eq:allowed_band_condition}
\end{equation}
whereas $|\Delta|>1$ defines the forbidden gaps. The band edges occur at $\Delta=+1$ or $-1$, corresponding respectively to $Q=0$ and $QL=\pm\pi$. The first Brillouin zone therefore has width $2\alpha$. The band equation obtained above coincides with that derived in reference \cite{GesztesyMacedoStreit1985}.%
\footnote{We thank Professor Fritz Gesztesy for bringing
Refs. \cite{GesztesyKirsch1985,GesztesyMacedoStreit1985} to our attention.} While the exact band equation agrees with the result of Ref.~\cite{GesztesyMacedoStreit1985}, the local connection-matrix construction developed here also determines the corresponding Bloch eigenvectors and provides the starting point for the physical analysis of the lattice. In particular, it allows the band edges to be related to the two scale-invariant spectra of the finite double-well problem and exposes two alternating families of gaps with distinct closing mechanisms. These results underlie the subsequent analysis of the symmetric and asymmetric limits, band-edge dispersions, effective masses, Dirac-like regimes, and normalized probability densities.

%While the exact band equation itself agrees with the result of Ref.~\cite{GesztesyMacedoStreit1985}, the purpose of the present work is not limited to its rederivation. The local connection-matrix formulation gives direct access to the Bloch eigenvectors and makes it possible to relate the band edges to the two scale-invariant spectra of the corresponding finite double-well problem. It also reveals two independent families of gaps and their distinct closing mechanisms, including the difference between the folded crossing of the symmetric lattice and the genuine linear contact of the asymmetric complementary-coupling lattice. These results provide the basis for the analysis of effective masses, Dirac-like crossover regimes, and normalized probability densities developed in the following sections.%Their treatment is based on self-adjoint extensions and the direct-integral decomposition of the periodic Hamiltonian, whereas the present derivation follows from local connection matrices and Bloch's condition. The agreement between the two approaches provides an independent consistency check of the matching prescription adopted here.% modulo reciprocal-lattice translations.

\subsection{Symmetric lattice}

The symmetric limit must be treated before completing the two-step transfer. When $\nu_s=\nu_c\equiv\nu$, translation by $\pi/(2\alpha)$ is already a symmetry, and the primitive period is reduced to $\ell=\pi/(2\alpha)$. Bloch's condition must consequently be imposed immediately after the first passage through a singularity. With the present conventions, the one-step transfer matrix is $\eta\mathcal C_{\mathrm{sym}}(\nu;\kappa)$, and hence
\begin{equation}
 \cos(Q\ell)=\frac{1}{2}\operatorname{tr}
 \left[\eta\mathcal C_{\mathrm{sym}}(\nu;\kappa)\right]
 =\frac{\cos(\pi\kappa/2)}{\sin(\pi\nu)}.
 \label{eq:symmetric_band_equation}
\end{equation}
Equation~\eqref{eq:symmetric_band_equation} has the same functional
form as the dispersion relation obtained in the WKB treatment of a
generic one-dimensional periodic potential~\cite{HeissenbergSagnotti2022}.
The correspondence is established by the formal identifications
\begin{equation}
 T=\sin(\pi\nu),
 \qquad
 \gamma(E)=\frac{\pi}{2}(\kappa-1) \ \ (\text{mod }2\pi),
\end{equation}
where $T$ is the WKB tunneling factor (the corresponding transmission probability being proportional to $T^2$) and
$\gamma(E)=\hbar^{-1}\int_{x_-}^{x_+}\sqrt{2m[E-V(x)]}\,dx-\pi/2$ ($x_\pm$ denoting the classical
turning points) is the phase accumulated within the classically allowed region. In the present model, however, the resulting band equation is
exact rather than semiclassical. This correspondence suggests that
$\sin(\pi\nu)$ measures the effective transmission through each inverse-square
singularity. In particular, its invariance under $\nu\mapsto1-\nu$ provides a
direct physical interpretation of the same discrete symmetry displayed by the
symmetric-lattice spectrum, while its maximum value at $\nu=1/2$ corresponds
to the free-particle limit, in which the singular potential disappears.

The corresponding first Brillouin zone has width $4\alpha$, twice that of the asymmetric case. Employing the doubled cell instead reproduces the same spectrum in a folded-zone representation. This distinction is essential when the symmetric and asymmetric dispersions are compared: the apparent meeting of two branches at the boundary of the reduced zone is a zone-folding effect, rather than the collision of two physically distinct bands. Its consequences for the gaps, band-edge dispersions, and nearly symmetric limit will be explored in the following sections. This result also agrees with the Bloch spectrum obtained in 
\cite{GesztesyKirsch1985}.
\begin{figure*}[t]
 \centering
 \includegraphics[width=\textwidth]{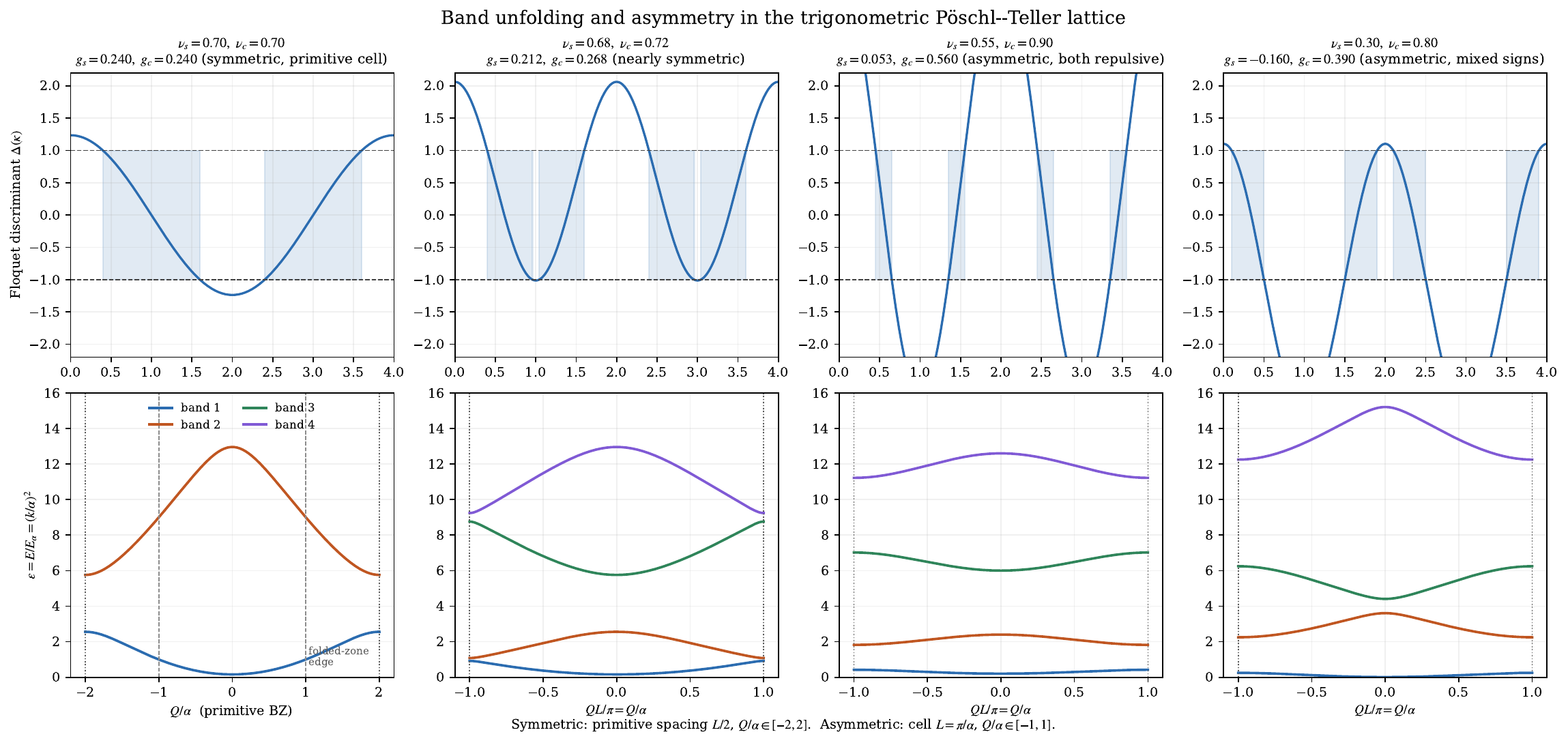}
 \caption{Discriminant given by equation~(\ref{discriminante_a})
 (upper panels) and the first four Bloch bands
 (lower panels) for representative lattices.  From left to right: the
 symmetric lattice, a nearly symmetric lattice, an asymmetric lattice with
 two repulsive singular couplings, and an asymmetric lattice with couplings
 of opposite signs.  The shaded intervals in the upper panels satisfy
 $|\Delta|\leq1$.  In the symmetric panel the shorter primitive period
 $\ell=\pi/(2\alpha)$ is used, whereas the asymmetric panels employ
 $L=\pi/\alpha$.}
 \label{fig:band_structure}
\end{figure*}

Figure~\ref{fig:band_structure} illustrates the equation~(\ref{discriminante_a}) and the first four energy bands for representative symmetric, nearly symmetric, and asymmetric lattices. It makes explicit both the opening of gaps when the two singular sublattices become inequivalent and the change in the Brillouin-zone width in the exactly
symmetric limit. The remainder of this article is devoted to a detailed study
of this band structure, including its spectral symmetries, band edges and gaps,
effective carrier masses, nearly symmetric dispersion, and the spatial
distribution of the corresponding Bloch states.

%%%%%%%%%%%%%%%%%%%%%%%%%%%%%%%%%%%%%%%%%%%%%%%%%%%%%%5
\section{Spectral properties of the Bloch bands}
\label{sec:spectral_properties}

The exact band equations derived in the preceding section make it
possible to determine the spectral properties of the lattice without
solving any additional differential equation. In this section, we first
obtain the band edges and show that the widths of successive bands and
gaps follow two alternating sequences. We then examine the symmetric
limit, in which the primitive cell is reduced by one half and pairs of
bands of the asymmetric description merge into a single band. Finally,
we briefly relate the band-edge energies to the discrete spectrum of the
symmetric double-well problem studied in Ref.~\cite{CamaraDaSilvaTPT}.

\subsection{Band edges and spectral gaps}
\label{subsec:band_edges}

For the asymmetric lattice, the exact dispersion relation is given by
equations~(\ref{eq:discriminant}) and~(\ref{discriminante_a}).
The band edges are determined by
\begin{equation}
 \Delta(\kappa)=\pm1
 \quad\Longleftrightarrow\quad
 QL=0,\pm\pi.
 \label{eq:band_edge_condition}
\end{equation}
The condition $\Delta=+1$, corresponding to $Q=0$, gives
\begin{equation}
 \cos(\pi\kappa)=-\cos\left[\pi(\nu_s+\nu_c)\right],
 \label{eq:edge_Q_zero}
\end{equation}
whereas $\Delta=-1$, corresponding to $QL=\pm\pi$, gives
\begin{equation}
 \cos(\pi\kappa)=-\cos\left[\pi(\nu_s-\nu_c)\right].
 \label{eq:edge_Q_pi}
\end{equation}

It is convenient to introduce the non-negative quantities
\begin{equation}
 a\equiv\left|1-(\nu_s+\nu_c)\right|,
 \qquad
 b\equiv1-\left|\nu_s-\nu_c\right|,
 \label{eq:edge_parameters}
\end{equation}
For $0<\nu_s,\nu_c<1$, these parameters satisfy
\begin{equation}
 0\leq a\leq b\leq1.
\end{equation}
The solutions of equations (\ref{eq:edge_Q_zero}) and
(\ref{eq:edge_Q_pi}) can then be arranged in increasing order as
\begin{eqnarray}
\kappa=
\left\{ \begin{array}{ll}
2j+a,       & Q=0,\\
2j+b,       & QL=\pm\pi,\\
2(j+1)-b,   & QL=\pm\pi,\\
2(j+1)-a,   & Q=0,
\end{array} \right.
\qquad j=0,1,2,\ldots
\label{eq:ordered_kappa_edges}
\end{eqnarray}
Thus, the first band begins at $Q=0$ and ends at $QL=\pm\pi$,
the second begins at $QL=\pm\pi$ and ends at $Q=0$, and this pattern repeats throughout the spectrum. Consequently, the allowed bands are
\begin{eqnarray}
&&\mathcal B_{2j+1}:\qquad
 2j+a\leq\kappa\leq2j+b,
 \label{eq:odd_bands}\\
&&\mathcal B_{2j+2}:\qquad
 2j+2-b\leq\kappa\leq2j+2-a.
 \label{eq:even_bands}
\end{eqnarray}
%In the terminology inherited from the symmetric double-well problem, the bands therefore alternate between even-to-odd and odd-to-even sequences.
Although the two kinds of bands have the same width in the variable $\kappa$,
\begin{equation}
 \Delta\kappa=b-a,
 \label{eq:kappa_band_width}
\end{equation}
their widths in energy are different. Defining $E=E_{\alpha}\kappa^2$, $E_{\alpha}=\hbar^2\alpha^2/2m$,
%\begin{equation}
% E=E_\alpha\kappa^2,
% \qquad
% E_\alpha\equiv\frac{\hbar^2\alpha^2}{2m}.
% \label{eq:energy_scale}
%\end{equation}
for the odd-numbered bands one obtains
\begin{equation}
 W_{2j+1}
 =
 E_\alpha
 \left[
 (2j+b)^2-(2j+a)^2
 \right]
 =
 E_\alpha(b-a)(4j+a+b),
 \label{eq:odd_band_width}
\end{equation}
whereas the even-numbered bands have width
\begin{equation}
 W_{2j+2}
 =
 E_\alpha
 \left[
 (2j+2-a)^2-(2j+2-b)^2
 \right]
 =
 E_\alpha(b-a)\left[4(j+1)-a-b\right].
 \label{eq:even_band_width}
\end{equation}
The distinction between the even-to-odd and odd-to-even bands is therefore absent in $\kappa$ space but becomes explicit in the physical energy spectrum. The forbidden intervals therefore form two alternating families. The gap separating $\mathcal B_{2j+1}$ from $\mathcal B_{2j+2}$ lies between
the two edges
\begin{equation}
 \kappa_{2j+1}^{(+)}=2j+b,
 \qquad
 \kappa_{2j+2}^{(-)}=2(j+1)-b,
 \label{eq:internal_gap_edges}
\end{equation}
and consequently has width
\begin{equation}
 \Delta\kappa_{2j+1}
 =2(1-b)=2|\nu_s-\nu_c|.
 \label{eq:internal_gap_kappa}
\end{equation}
In energy variables, this gives
\begin{eqnarray}
 \Delta E_{2j+1}
 =
 E_\alpha
 \left[
 (2(j+1)-b)^2-(2j+b)^2
 \right]
 =4E_\alpha(2j+1)(1-b)
 =4E_\alpha(2j+1)|\nu_s-\nu_c|.
 \label{eq:asymmetry_gap}
\end{eqnarray}
This gap is generated entirely by the inequivalence of the two singular sublattices and will therefore be referred to as the \emph{intrapair} or \emph{asymmetry gap}.

The second type of forbidden interval separates $\mathcal B_{2j+2}$ from $\mathcal B_{2j+3}$. Its edges are
\begin{equation}
 \kappa_{2j+2}^{(+)}=2j+2-a,
 \qquad
 \kappa_{2j+3}^{(-)}=2j+2+a,
 \label{eq:external_gap_edges}
\end{equation}
so that
\begin{equation}
 \Delta\kappa_{2j+2}=2a
 =2|1-\nu_s-\nu_c|.
 \label{eq:external_gap_kappa}
\end{equation}
The corresponding energy gap is
\begin{eqnarray}
 \Delta E_{2j+2}
 &=
 E_\alpha
 \left[
 (2j+2+a)^2-(2j+2-a)^2
 \right]
 \nonumber\\
 &=
 8E_\alpha(j+1)a
 =8E_{\alpha}(j+1)|1-(\nu_s+\nu_c)|.
 \label{eq:interpair_gap}
\end{eqnarray}
We shall refer to this second family as the \emph{interpair gaps}.

The distinction between these two families is essential. %Introducing the
%mean coupling
%\begin{equation}
% \bar{\nu}=\frac{\nu_s+\nu_c}{2},
%\end{equation}
The two gap families are controlled independently by
$|\nu_s-\nu_c|$ and the mean coupling $\bar{\nu}=(\nu_s+\nu_c)/2$, respectively. Consequently, the intrapair gaps shrink continuously and vanish whenever $|\nu_s-\nu_c|\rightarrow0$, irrespective of how $\bar{\nu}$ varies along this limit. Likewise, the interpair gaps vanish whenever $\bar{\nu}\rightarrow\frac12$, irrespective of the behavior of $|\nu_s-\nu_c|$. In short:
\begin{eqnarray}
 &&|\nu_s-\nu_c|\rightarrow0
 \quad\Longrightarrow\quad
 \Delta E_{2j+1}\rightarrow0,\nonumber\\
 &&\bar{\nu}\rightarrow\frac12
 \quad\Longrightarrow\quad
 \Delta E_{2j+2}\rightarrow0.
\end{eqnarray}
The first limit is accompanied by a reduction of the primitive period and is examined in the following subsection. The second leaves the asymmetric primitive cell unchanged whenever $|\nu_s-\nu_c|\neq0$; its main physical consequence emerges from the local dispersion relation and will be discussed in the subsequent analysis of the band-edge dispersions.
%The distinction between these two families is essential. At fixed mean coupling $\bar{\nu}=(\nu_s+\nu_c)/2$, the parameter $a=|1-2\bar{\nu}|$ remains unchanged as the difference $|\nu_s-\nu_c|$ is reduced. Consequently, the intrapair gaps shrink continuously and vanish in the symmetric limit $|\nu_s-\nu_c|\to0$, whereas the interpair gaps remain finite unless $\bar{\nu}=1/2$. Conversely, the interpair gaps vanish when $a=0$, or equivalently when $\nu_s+\nu_c=1$, even though the intrapair gaps generally remain open. The model therefore possesses two distinct gap-closing conditions: $\nu_s=\nu_c$ closes $\Delta E_{2j+1}$, while $\nu_s+\nu_c=1$ closes $\Delta E_{2j+2}$. The first limit is accompanied by a reduction of the primitive period and is examined in the following subsection. The second leaves the asymmetric primitive cell unchanged; its main physical consequence emerges from the local dispersion relation and will be discussed in the subsequent analysis of the band-edge dispersions.

\subsection{The symmetric lattice and Brillouin-zone unfolding}
\label{subsec:symmetric_limit}

When $\nu_s=\nu_c=
\bar\nu$, one has
\begin{equation}
 b=1,
 \qquad
 a=|1-2\bar\nu|.
 \label{eq:symmetric_edge_parameters}
\end{equation}
The mechanism by which the intrapair gap disappears can be seen directly
from its two limiting edges. For nonzero asymmetry, the upper edge of
$\mathcal B_{2j+1}$ and the lower edge of $\mathcal B_{2j+2}$ are
located at
\begin{equation}
 \kappa=2j+1-|\nu_s-\nu_c|
 \qquad\text{and}\qquad
 \kappa=2j+1+|\nu_s-\nu_c|,
\end{equation}
respectively. As $|\nu_s-\nu_c|\to0$, these two edges approach the
common value
\begin{equation}
 \kappa=2j+1.
\end{equation}
The forbidden interval between them therefore contracts continuously
to a single point. At exact symmetry, the two asymmetric branches become
the two folded portions of the same symmetric band,
\begin{equation}
 [\,2j+a,\,2j+1\,]
 \cup
 [\,2j+1,\,2j+2-a\,]
 =
 [\,2j+a,\,2j+2-a\,].
 \label{eq:symmetric_band_merging}
\end{equation}

The closing of the intrapair gap is therefore accompanied by a change
in the translational symmetry of the lattice. For
$|\nu_s-\nu_c|\neq0$, the two
singular sites are inequivalent and the primitive period is
$L=\pi/\alpha$. At $\nu_s=\nu_c$, they become equivalent, and
translation by only $\ell=L/2=\pi/(2\alpha)$ is sufficient to
reproduce the potential.
The Brillouin zone consequently doubles in width. What appears in the
asymmetric reduced-zone description as two bands touching at
$\kappa=2j+1$ must therefore be unfolded into a single band of the
symmetric lattice.
This mechanism is analogous to the undimerized limit of the
Su--Schrieffer--Heeger chain, in which equality of the alternating hopping
amplitudes closes the dimerization gap and simultaneously reduces the
primitive cell; the crossing seen in the doubled-cell description is then
the folded representation of the uniform-chain spectrum~\cite{SuSchriefferHeeger1979}.

By contrast, the neighboring interpair gap remains bounded by
\begin{equation}
 \kappa=2j+2-a
 \qquad\text{and}\qquad
 \kappa=2j+2+a.
\end{equation}
Since $a=|1-2\bar{\nu}|$ does not vanish when the asymmetry is removed
at fixed $\bar{\nu}$, this forbidden interval survives in the symmetric
lattice. Hence, the symmetric limit does not eliminate the complete gap
structure: it removes only the gaps that separate the two folded
portions of each symmetric band.

\begin{figure*}[t]
 \centering
 \includegraphics[scale=0.55]{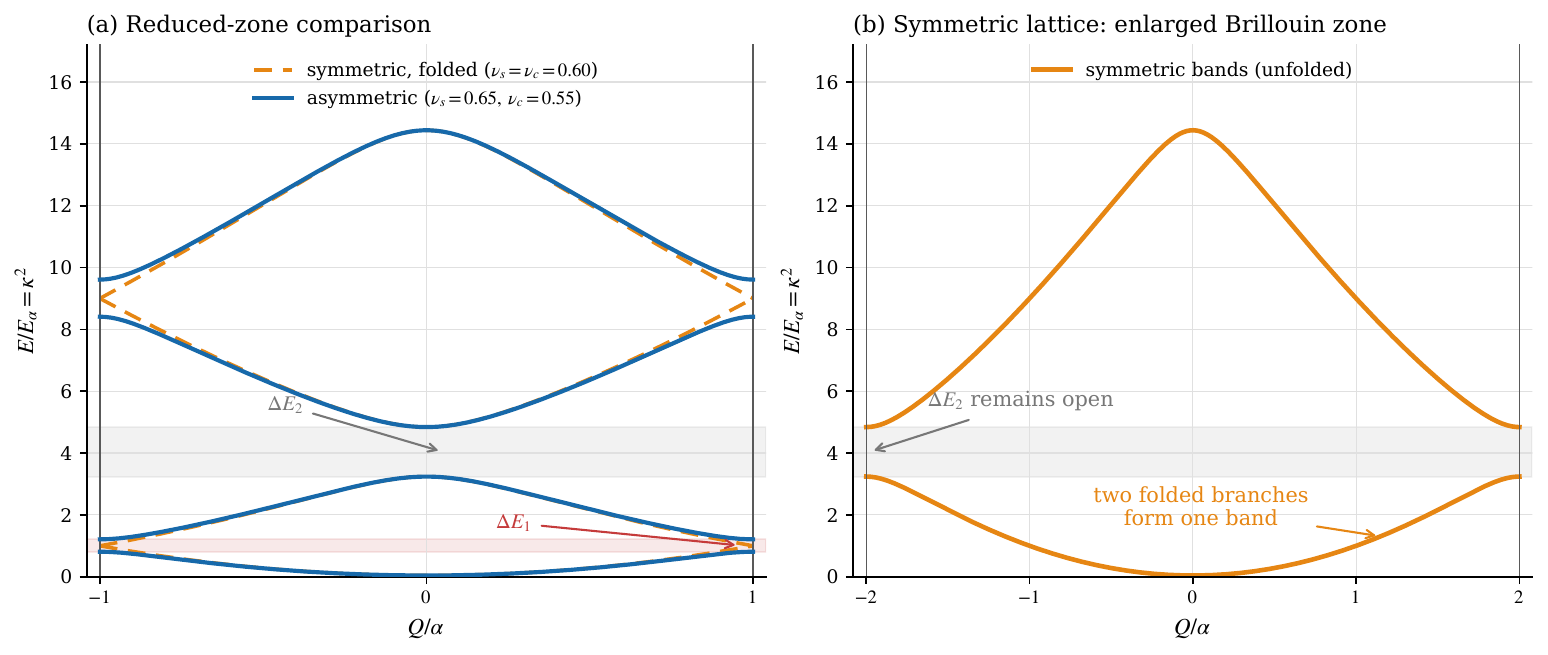}
 \caption{Comparison between the asymmetric and symmetric band structures at fixed mean coupling $\bar{\nu}$. (a) Reduced-zone representation associated with the asymmetric primitive period $L=\pi/\alpha$. The solid curves denote the asymmetric bands, whereas the dashed curves show the folded symmetric bands. The lower shaded region marks the asymmetry gap $\Delta E_1$, while the upper one marks
the neighboring gap $\Delta E_2$. (b) Extended-zone representation showing that, when $\nu_s=\nu_c$, the two folded branches form a single band in the enlarged Brillouin zone. The intrapair gaps close, whereas the interpair gaps remain finite.}
 \label{fig:band_limite}
\end{figure*}

This change of spectral organization is illustrated in
figure~\ref{fig:band_limite}. In the reduced-zone representation, the
asymmetry gap $\Delta E_1$ closes continuously as
$|\nu_s-\nu_c|\to0$, whereas the neighboring gap $\Delta E_2$ remains
open. Once the symmetric lattice is described using its true primitive
cell, the two touching reduced-zone branches unfold into a single band
over the enlarged Brillouin zone.

\subsection{Relation to the discrete double-well spectrum}
\label{subsec:double_well_correspondence}

The band-edge energies have a direct interpretation in terms of the
discrete spectrum of the symmetric double-well problem constructed in
Ref.~\cite{CamaraDaSilvaTPT}. When both couplings lie in the region
$0<\nu_s,\nu_c<1$, two scale-invariant choices of boundary conditions
are available at the singular endpoints. Neither choice introduces an
additional length scale, and both therefore preserve the local
conformal symmetry of the inverse-square singularities.

For the first scale-invariant choice, the even- and odd-parity spectra
are
\begin{equation}
 \frac{k_n^{(\mathrm{e},1)}}{\alpha}
 =2n+1-\nu_s+\nu_c,
 \qquad
 \frac{k_n^{(\mathrm{o},1)}}{\alpha}
 =2n+1+\nu_s+\nu_c,
 \qquad n=0,1,2,\ldots.
 \label{eq:double_well_spectrum_1}
\end{equation}
For the second scale-invariant choice, they are
\begin{equation}
 \begin{split}
 \frac{k_0^{(\mathrm{e},2)}}{\alpha}
 &=|1-\nu_s-\nu_c|,
 \\[2mm]
 \frac{k_n^{(\mathrm{e},2)}}{\alpha}
 &=2n+1-\nu_s-\nu_c,
 \qquad n=1,2,\ldots,
 \\[2mm]
 \frac{k_n^{(\mathrm{o},2)}}{\alpha}
 &=2n+1+\nu_s-\nu_c,
 \qquad n=0,1,2,\ldots.
 \end{split}
 \label{eq:double_well_spectrum_2}
\end{equation}
Here, $\mathrm{e}$ and $\mathrm{o}$ denote the even- and odd-parity
sectors of the symmetric double well, while the labels $1$ and $2$
distinguish the two scale-invariant boundary conditions.

The union of the two spectra,
\begin{equation}
 \sigma_{\mathrm{fp}}
 =
 \sigma_1\cup\sigma_2,
 \label{eq:combined_fixed_point_spectrum}
\end{equation}
can be arranged in increasing order by using the parameters introduced
in equation~(\ref{eq:edge_parameters}). The resulting sequence is
\begin{equation}
 \kappa=
 2j+a,\quad
 2j+b,\quad
 2(j+1)-b,\quad
 2(j+1)-a,
 \qquad j=0,1,2,\ldots.
 \label{eq:ordered_double_well_spectrum}
\end{equation}
This is precisely the sequence of Bloch-band edges obtained in
equation~(\ref{eq:ordered_kappa_edges}). Therefore, the complete set of
band edges is not reproduced by either scale-invariant realization
separately; it is obtained from the union of their discrete spectra.

This correspondence has the standard physical interpretation expected
when a finite well is periodically repeated. The discrete levels of the
isolated system broaden into Bloch bands as neighboring wells are
coupled, while the spectra selected by the scale-invariant boundary
conditions determine their limiting energies. The agreement provides
both a physical interpretation of the exact band edges and an
independent consistency check of the periodic construction.

%%%%%%%%%%%%%%%%%%%%%%%%%%%%%%%%%%%%%%%%%%%%%%%%%%%%%

\section{Band-edge dispersion and effective carrier masses}
\label{sec:band_edge_dispersion}

The exact band equation also determines the local dynamics of the Bloch
states near the band edges.  Away from a gap-closing condition, every edge
has the usual quadratic dispersion and can be characterized by an effective
mass.  When one of the two alternating gap families becomes small, however,
the quadratic region shrinks and crosses over to a Dirac-like dispersion.
The two possible gap-closing mechanisms must be distinguished: the nearly
symmetric limit, $|\nu_s-\nu_c|\ll1$, closes the gaps at the boundary of the
Brillouin zone, whereas $|1-\nu_s-\nu_c|\ll1$ closes the complementary gaps
at its center.

For later convenience, we define
\begin{equation}
 \sigma\equiv
 \sin(\pi\nu_s)\sin(\pi\nu_c)>0
 \label{eq:sigma_definition}
\end{equation}
and write the two edges surrounding a given gap as
\begin{equation}
 \kappa_{h,N}=N-\eta_N,
 \qquad
 \kappa_{e,N}=N+\eta_N,
 \label{eq:kappa_edge_unified}
\end{equation}
where
\begin{equation}
 \begin{array}{lll}
  N=2j+1, & \eta_N=|\nu_s-\nu_c|,
          & Q_{\rm c}L=\pm\pi,\\[1mm]
  N=2j+2, & \eta_N=|1-\nu_s-\nu_c|,
          & Q_{\rm c}=0.
 \end{array}
 \label{eq:gap_families_unified}
\end{equation}
Here $\kappa_{h,N}$ is the upper edge of the lower band and
$\kappa_{e,N}$ is the lower edge of the upper band.  We also introduce the
momentum measured from the corresponding contact point,
\begin{equation}
 p\equiv\hbar(Q-Q_{\rm c}).
 \label{eq:local_momentum}
\end{equation}

\subsection{Non-critical band edges}
\label{subsec:generic_effective_masses}

We first assume that both $|\nu_s-\nu_c|$ and
$|1-\nu_s-\nu_c|$ are nonzero. Expanding
equation~(\ref{eq:discriminant}) around $Q=Q_{\rm c}$ gives
\begin{eqnarray}
 E_h(p)&=E_\alpha(N-\eta_N)^2
        -\frac{p^2}{2m^*_{h,N}}+O(p^4),
 \label{eq:hole_parabolic_dispersion}\\
 E_e(p)&=E_\alpha(N+\eta_N)^2
        +\frac{p^2}{2m^*_{e,N}}+O(p^4).
 \label{eq:electron_parabolic_dispersion}
\end{eqnarray}
The positive electron and hole effective masses are
\begin{eqnarray}
 m^*_{e,N}
 =\frac{m\sin(\pi\eta_N)}
 {\pi\sigma(N+\eta_N)}, \ \ 
 m^*_{h,N}=\frac{m\sin(\pi\eta_N)}
 {\pi\sigma(N-\eta_N)}.
 \label{eq:hole_effective_mass}
\end{eqnarray}
The minus sign in equation~(\ref{eq:hole_parabolic_dispersion}) expresses the
negative curvature at the top of the lower band.  The corresponding missing
particle is described as a positively charged hole with the positive mass
$m^*_{h,N}$.  Thus the present one-particle problem does not contain two
independent charge species; the electron--hole terminology refers to the
quasiparticle interpretation of a partially filled band.

\begin{figure*}[t]
 \centering
 \includegraphics[width=0.92\textwidth]{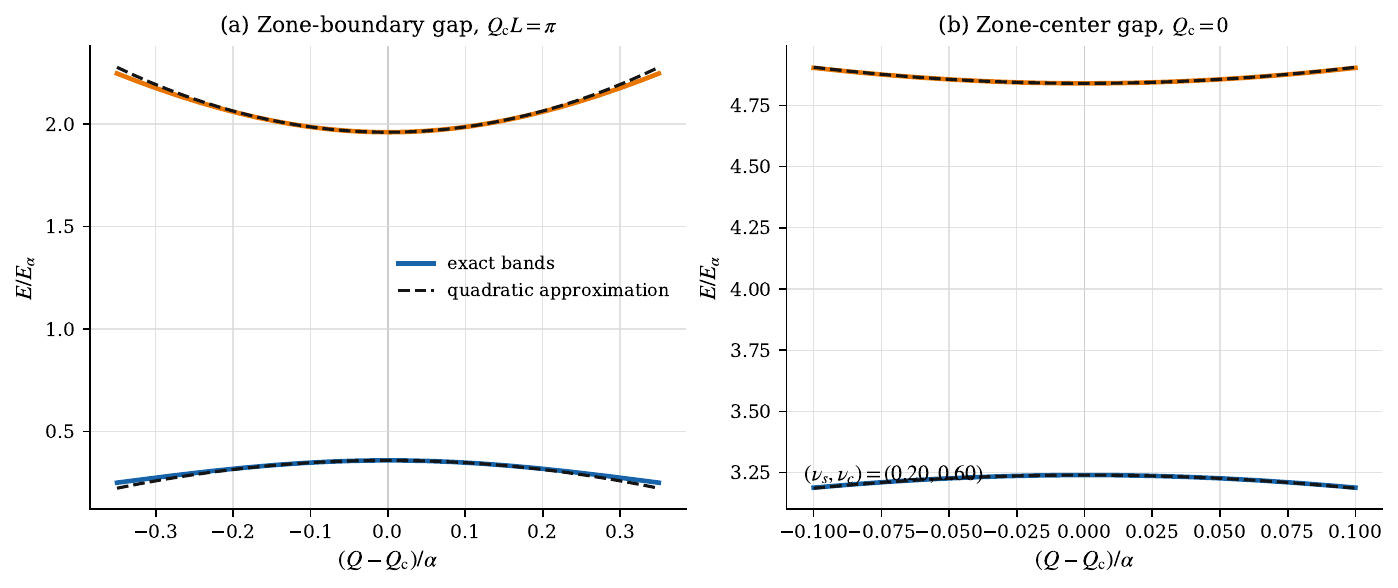}
 \caption{Exact dispersion (solid curves) and quadratic band-edge
 approximations (dashed curves) for $(\nu_s,\nu_c)=(0.20,0.60)$.
 (a) The gap centered at the boundary of the Brillouin zone, for which
 $N=1$ and $\eta_N=|\nu_s-\nu_c|$.
 (b) The next gap, centered at $Q=0$, for which
 $N=2$ and $\eta_N=|1-\nu_s-\nu_c|$.}
 \label{fig:generic_edge_dispersion}
\end{figure*}

The exact bands and their quadratic approximations are compared in
figure~\ref{fig:generic_edge_dispersion}.  The approximation is tangent to
the exact result at each edge and remains accurate within a finite
neighborhood whenever $\eta_N$ is not small.

\subsection{Universal form near a closing gap}
\label{subsec:universal_dirac_form}

The preceding expansion is not uniform when $\eta_N\rightarrow0$, because
both effective masses vanish and the quadratic neighborhood collapses.  To
obtain the appropriate approximation, $\kappa-N$ and $Q-Q_{\rm c}$ must be
expanded simultaneously while $\eta_N$ is retained.  The exact band equation
then gives
\begin{equation}
 (\kappa-N)^2
 =\eta_N^2+
 \sigma\left(\frac{p}{\hbar\alpha}\right)^2
 +O\!\left(\eta_N^4,p^4,\eta_N^2p^2\right).
 \label{eq:kappa_hyperbola}
\end{equation}
Since $E=E_\alpha\kappa^2$, equation~(\ref{eq:kappa_hyperbola}) yields
\begin{equation}
 E_\pm(p)=E_{D,N}
 +\frac{\sigma p^2}{2m}
 \pm\sqrt{\left(m^*_{D,N}c_N^2\right)^2+c_N^2p^2}
 +O\!\left(\eta_N^3,p^3\right),
 \label{eq:massive_dirac_dispersion}
\end{equation}
where
\begin{eqnarray}
 E_{D,N}=E_\alpha(N^2+\eta_N^2),\ \  c_N&=\frac{\hbar\alpha N}{m}\sqrt{\sigma}, \ \ m^*_{D,N}&=\frac{m\eta_N}{N\sigma}.
 \label{eq:dirac_mass}
\end{eqnarray}
Equation~(\ref{eq:massive_dirac_dispersion}) also contains the small
particle--hole-asymmetric correction $\sigma p^2/(2m)$, as expected for an
effective description derived from an underlying nonrelativistic theory.
Neglecting this subleading contribution gives the standard one-dimensional massive ``Dirac-like'' dispersion~\cite{GarreauZehnle2017}. Since the quadratic term is common to both branches, it cancels from their energy difference,
\begin{equation}
 E_{+}(p)-E_-(p)
 \approx 2\sqrt{\left(m^*_{D,N}c_N^2\right)^2+c_N^2p^2}.
 \label{eq:dirac_band_splitting}
\end{equation}
In particular, at the band edges ($p=0$),
\begin{equation}
 m^*_{D,N}c_N^2=2E_\alpha N\eta_N
 =\frac{1}{2}\left(E_{+}(0)-E_-(0)\right),
 \label{eq:rest_energy_half_gap}
\end{equation}
so that the rest-energy scale is exactly one half of the corresponding gap.

The crossover momentum is set by
\begin{equation}
 \bar p_{N}=m^*_{D,N}c_N
 =\frac{\hbar\alpha\eta_N}{\sqrt{\sigma}}.
 \label{eq:crossover_momentum}
\end{equation}
For $|p|\ll \bar p_{N}$, the two branches reduce again to parabolic
dispersions. To leading order in $\eta_N$,
\begin{equation}
 \frac{1}{m^*_{e,N}}
 =\frac{1}{m^*_{D,N}}+\frac{\sigma}{m}
 +O\!\left(\frac{\eta_N}{m}\right),
 \qquad
 \frac{1}{m^*_{h,N}}
 =\frac{1}{m^*_{D,N}}-\frac{\sigma}{m}
 +O\!\left(\frac{\eta_N}{m}\right).
 \label{eq:edge_masses_dirac_limit}
\end{equation}
By contrast, for $\bar p_{N}\ll|p|\ll\hbar\alpha$, the dispersion is
approximately linear. Therefore, for every nonzero gap the immediate
neighborhood of a band edge remains nonrelativistic; the Dirac-like regime
emerges outside this progressively narrow region as the gap closes.

\subsection{Nearly symmetric lattice}
\label{subsec:nearly_symmetric_dispersion}

\begin{figure*}[t]
 \centering
 \includegraphics[width=0.92\textwidth]{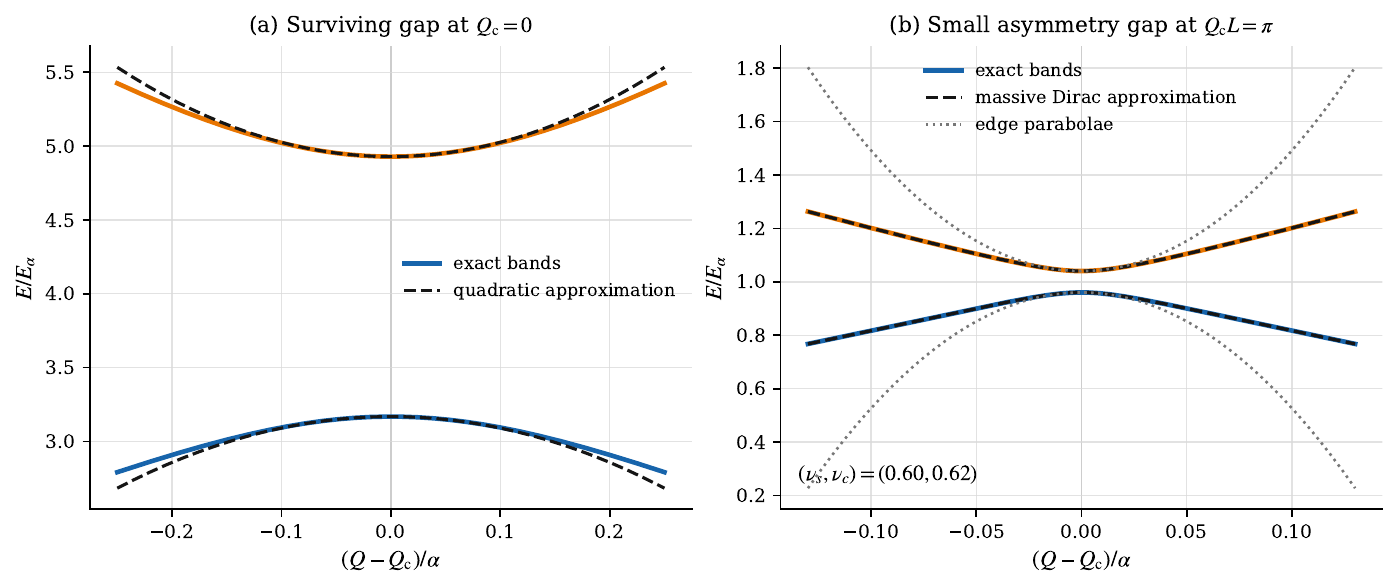}
 \caption{Band-edge dispersion for the nearly symmetric lattice
 $(\nu_s,\nu_c)=(0.60,0.62)$.  (a) The finite gap at $Q=0$ retains the
 usual quadratic behavior.  (b) Near the small asymmetry gap at the zone
 boundary, the massive Dirac approximation (dashed) follows the exact bands
 well beyond the narrow region described by the edge parabolae (dotted).}
 \label{fig:nearly_symmetric_crossover}
\end{figure*}

Consider first
\begin{equation}
 |\nu_s-\nu_c|\ll1,
 \qquad
 |1-\nu_s-\nu_c|\not\simeq0.
 \label{eq:nearly_symmetric_condition}
\end{equation}
The gaps at $Q_{\rm c}L=\pm\pi$, corresponding to odd $N=2j+1$,
become small. Their local dispersion is given by
equation~(\ref{eq:massive_dirac_dispersion}) with
$\eta_N=|\nu_s-\nu_c|$ and
\begin{equation}
 m^*_{D,N}=\frac{m|\nu_s-\nu_c|}{N\sigma}.
 \label{eq:near_symmetric_dirac_mass}
\end{equation}
The gaps centered at $Q=0$, in contrast, retain
$\eta_N=|1-\nu_s-\nu_c|$ and therefore remain generically quadratic.  This
coexistence is illustrated in figure~\ref{fig:nearly_symmetric_crossover}.
The narrow avoided crossing is described by the massive Dirac expression
over a much wider interval than by either edge parabola, thereby displaying
the crossover between the quadratic and approximately linear regimes.

At $\nu_s=\nu_c$ the Dirac mass associated with the odd-$N$ gaps vanishes. One must nevertheless recall that the primitive period is then reduced from $\pi/\alpha$ to $\pi/(2\alpha)$.  Consequently, the resulting crossing belongs to the folded-zone representation of the symmetric lattice; after unfolding the Brillouin zone, the two branches form a single band.  The even-$N$ gaps remain as the ordinary gaps of the symmetric lattice.
%%%%%%%%%%%%%%%%%%%%%%%%%%%%%%%%%%

\subsection{Complementary couplings and a massless Dirac-like contact}
\label{subsec:complementary_dirac_contact}

The second gap-closing mechanism occurs when
\begin{equation}
 |1-\nu_s-\nu_c|\ll1,
 \qquad
 |\nu_s-\nu_c|\not\simeq0.
 \label{eq:near_complementary_condition}
\end{equation}
Now it is the even-$N$ family of gaps at $Q_{\rm c}=0$ that acquires the Dirac form, with
\begin{equation}
 m^*_{D,N}=\frac{m|1-\nu_s-\nu_c|}{N\sigma},
 \qquad N=2j+2.
 \label{eq:complementary_dirac_mass}
\end{equation}
The odd-$N$ gaps at the boundary of the Brillouin zone remain finite and retain their conventional quadratic edges.

\begin{figure*}[t]
 \centering
 \includegraphics[width=0.92\textwidth]{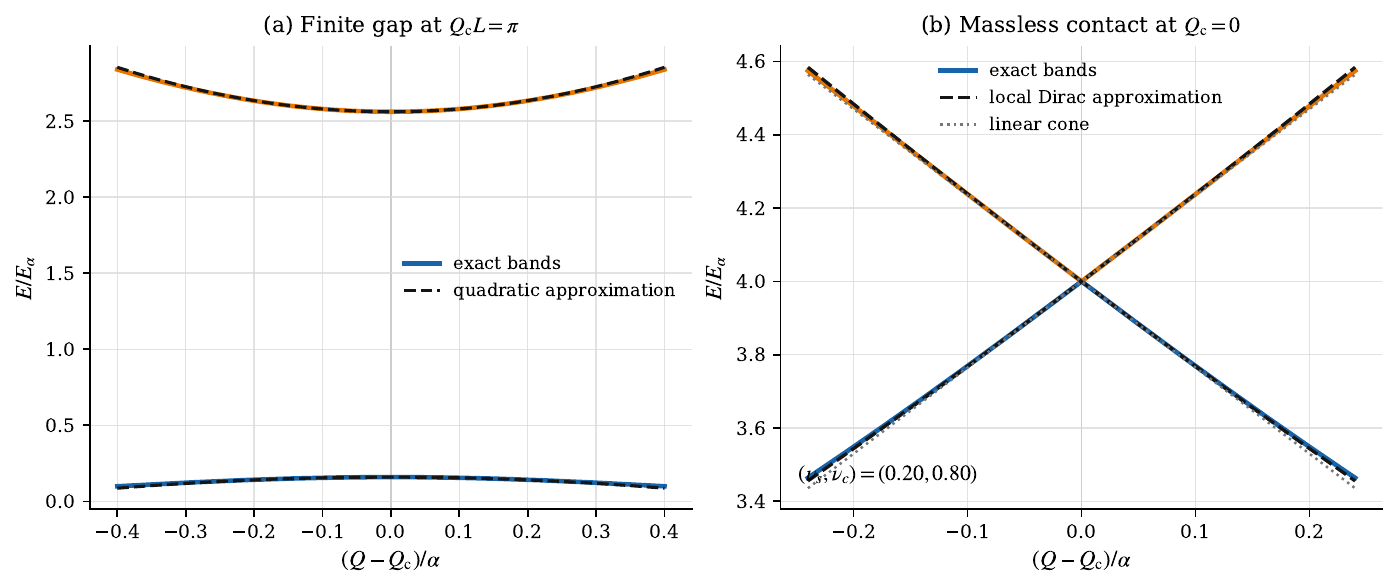}
 \caption{Dispersion on the complementary-coupling line
 $(\nu_s,\nu_c)=(0.20,0.80)$.  (a) The zone-boundary asymmetry gap remains
 finite and has ordinary quadratic edges.  (b) The even-$N$ gap closes at
 $Q=0$, producing a massless Dirac-like contact.  The dashed curves include
 the common quadratic correction in equation~(\ref{eq:massless_dirac_dispersion}),
 while the dotted curves show the strictly linear cone.}
 \label{fig:complementary_dirac_contact}
\end{figure*}

\begin{figure*}[t]
 \centering
\includegraphics[width=0.7\textwidth]
 {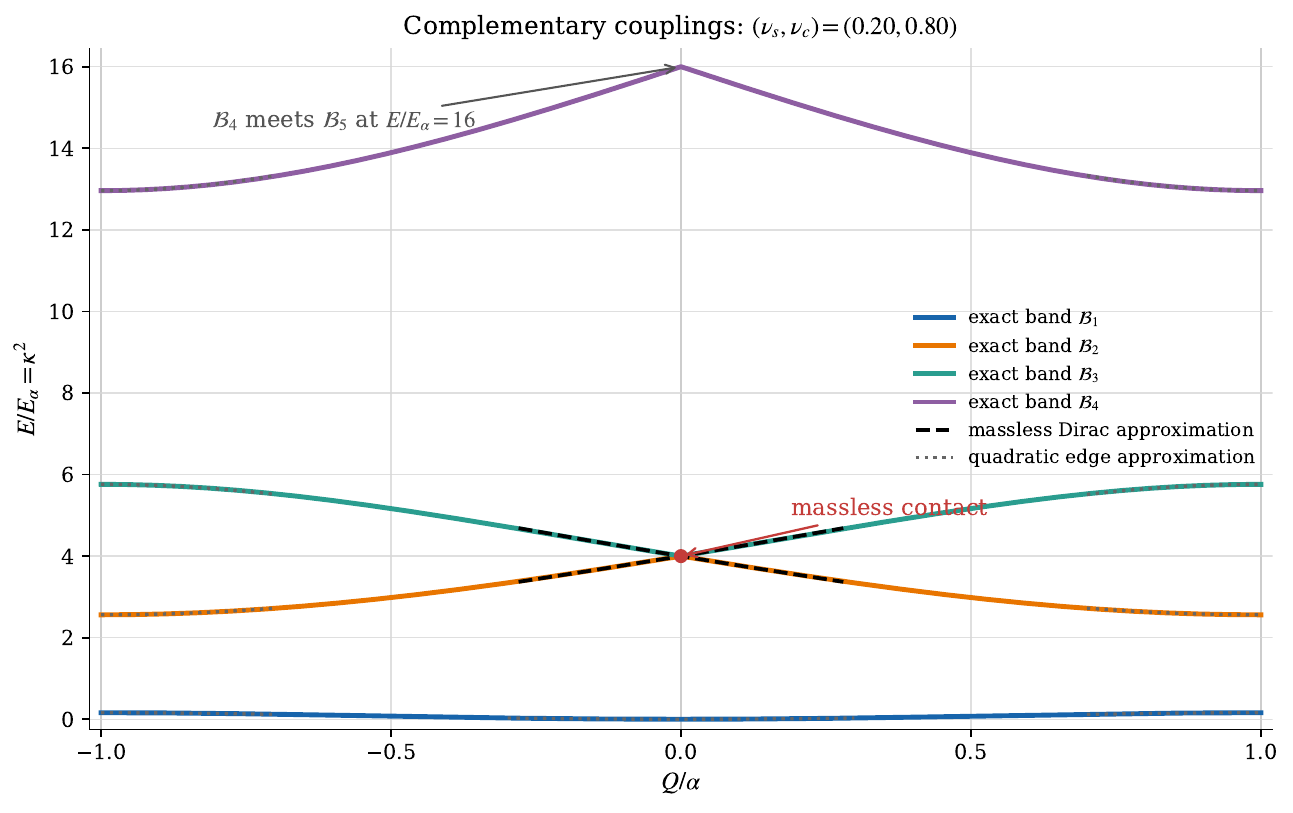}
 \caption{First four exact Bloch bands in the first Brillouin zone for the
 complementary couplings $(\nu_s,\nu_c)=(0.20,0.80)$. The dashed curves
 show the massless Dirac approximation around the contact between
 $\mathcal B_2$ and $\mathcal B_3$ at $Q=0$, whereas the dotted curves show
 the quadratic approximations near the regular band edges at
 $Q/\alpha=0,\pm1$. The fourth band meets the fifth band (not shown) at
 $E/E_\alpha=16$.}
 \label{fig:complementary_full_bands}
\end{figure*}

On the critical-coupling line,
\begin{equation}
 \nu_s+\nu_c=1,
 \label{eq:complementary_line}
\end{equation}
one has $m^*_{D,N}=0$, and equation~(\ref{eq:massive_dirac_dispersion})
reduces to
\begin{equation}
 E_\pm(p)-E_\alpha N^2=
 \frac{\sigma p^2}{2m}\pm c_N|p|+O(p^3).
 \label{eq:massless_dirac_dispersion}
\end{equation}
Thus the two bands touch linearly at $Q=0$, with slopes $\pm c_N$.
In contrast to the crossing obtained in the exactly symmetric lattice, this
contact does not result from a reduction of the primitive cell: for
$\nu_s\neq\nu_c$, the period remains $L$. It therefore represents a genuine
massless Dirac-like contact---a Dirac cone---in the original Brillouin zone.
Figure~\ref{fig:complementary_dirac_contact} contrasts the surviving
quadratic gap with this massless contact. While that figure focuses on the
local dispersion near the two types of band edges,
figure~\ref{fig:complementary_full_bands} shows how they coexist throughout
the first Brillouin zone. The second and third bands form a massless
Dirac-like contact at $Q=0$, whereas the finite gaps at $Q/\alpha=\pm1$
retain their conventional quadratic edges. The fourth band reaches the
lower branch of the next Dirac-like contact at $E/E_\alpha=16$; its upper
branch belongs to the fifth band and is not displayed.

\section{Bloch-state probability densities}
\label{sec:bloch_densities}
%%%%%%%%%%%%%%%%%%%%%%%%%%%%%%%%%%%%%%%%%%%
%%%%%%%%%%%%%%%%%%%%%%%%%%%%%%%%%%%%%%%%%%%%%%%%%%%%%%%%%%
The dispersion relation determines the allowed energies, but it does not by itself specify how a Bloch state is distributed within the primitive cell. This information is contained in the corresponding eigenvector of the transfer matrix.  In the present model, the eigenvector fixes the relative weight of the linearly independent solutions and therefore completes the determination of the wave function, apart from an overall normalization.%  We emphasize that the quantity considered in this section is the probability density of a single-electron Bloch state, rather than the many-electron density of a partially filled band.

\subsection{Bloch eigenvectors and the coefficient ratio}
\label{subsec:bloch_eigenvectors}

%At a singularity of type $s$, introduce the coefficient vector
%\begin{equation}
 %\mathbf u_s(k)=
 %\begin{pmatrix}
 % \alpha^{-\nu_s}A_0(k)\\
 % \alpha^{\nu_s}B_0(k)
 %\end{pmatrix}
%\end{equation}
Solving the Bloch condition, equation~(\ref{eq:bloch_eigenvalue}), gives the dimensionless ratio
\begin{equation}
 r_s(k,Q)\equiv
 \alpha^{2\nu_s}\frac{B_0(k)}{A_0(k)}.
 \label{eq:rs_definition}
\end{equation}
%Bloch's condition, equation~(\ref{eq:bloch_eigenvalue}), gives
%\begin{equation}
% \mathcal T(\kappa)\mathbf u_s(k)
% =e^{iQL}\mathbf u_s(k),
 %\qquad
 %\kappa=\frac{k}{\alpha},
 %\qquad
 %L=\frac{\pi}{\alpha}.
 %\label{eq:bloch_eigenvector_density_section}
%\end{equation}
%For
%\begin{equation}
% \mathcal T=
% \begin{pmatrix}
%  \mathcal T_{11}&\mathcal T_{12}\\
% \mathcal T_{21}&\mathcal T_{22}
% \end{pmatrix},
%\end{equation}
%the ratio follows from either row of the eigenvalue equation:
\begin{equation}
 r_s(k,Q)=
 \frac{e^{iQL}-\mathcal T_{11}}{\mathcal T_{12}}
 =\frac{\mathcal T_{21}}{e^{iQL}-\mathcal T_{22}}.
 \label{eq:rs_transfer_entries}
\end{equation}
Using $e^{\pm iQL}=\Delta(\kappa)\pm i\sqrt{1-\Delta^2(\kappa)}$, as follows from equation~(\ref{discriminante_a}), equation~(\ref{eq:rs_transfer_entries}) can be written as
\begin{equation}
 r_s^{(\pm)}=
 \frac{\mathcal T_{22}-\mathcal T_{11}}{2\mathcal T_{12}}
 \pm i\frac{\sqrt{1-\Delta^2}}{\mathcal T_{12}}.
 \label{eq:rs_general}
\end{equation}
The two signs correspond to the counterpropagating, time-reversed states with momenta $Q$ and $-Q$. Because the transfer matrix is real inside an allowed band, $r_s^{(-)}=[r_s^{(+)}]^*$, and the two states have the same probability density.

The ratio at the neighboring $c$ singularity is obtained without solving a
second eigenvalue problem.  If
\begin{equation}
 \mathcal C_{sc}=
 \begin{pmatrix}
  C_{11}^{sc}&C_{12}^{sc}\\
  C_{21}^{sc}&C_{22}^{sc}
 \end{pmatrix}
\end{equation}
is the connection matrix from the $s$ site to the $c$ site, then
\begin{equation}
 r_c^-(k,Q)\equiv
 \alpha^{2\nu_c}\frac{B_1(k)}{A_1(k)}
 =
 \frac{C_{21}^{sc}+C_{22}^{sc}r_s}
      {C_{11}^{sc}+C_{12}^{sc}r_s}.
 \label{eq:rc_from_rs}
\end{equation}
%The matching prescription changes the sign of the first local coefficient,
%so the ratio immediately to the right of the same singularity is
%\begin{equation}
% r_c^+(k,Q)=
% \alpha^{2\nu_c}\frac{\widetilde B_1(k)}{\widetilde A_1(k)}
% =-r_c^-(k,Q).
% \label{eq:rc_matching_sign}
%\end{equation}
Equations~(\ref{eq:rs_general}) and~(\ref{eq:rc_from_rs}), together with the cell solutions of section~\ref{sec:cell_solution}, determine the complete
Bloch wave function up to an overall constant, since the matching conditions propagate the ratios $B_0/A_0$ and $B_1/A_1$ to every cell of the lattice. Explicit expressions for the connection and transfer matrices are collected in appendix~\ref{app:transfer_eigenvectors}.

\subsection{Conformal band-edge states}
\label{subsec:conformal_band_edges}

The coefficient ratios also clarify the relation between the Bloch-band
edges and the scale-invariant spectra of the isolated double well.  A generic
linear combination of the two local behaviors contains the dimensionful
ratio $B_i/A_i$ and therefore introduces a local scale.  At a fixed-point
boundary condition, by contrast, one coefficient vanishes:
\begin{equation}
 \begin{array}{lll}
  r_i=0      &\Longleftrightarrow&B_i=0,
  \qquad \psi\sim |x-x_i|^{\frac12+\nu_i},\\[1mm]
  r_i=\infty &\Longleftrightarrow&A_i=0,
  \qquad \psi\sim |x-x_i|^{\frac12-\nu_i}.
 \end{array}
 \label{eq:pure_asymptotic_behaviors}
\end{equation}
The scale associated with the ratio $B_i/A_i$ then disappears and the wave
function has a pure power-law asymptotic form. These are precisely the two boundary conditions that preserve the asymptotic conformal symmetry of the inverse-square interaction. These special values arise from the poles of the Gamma functions entering the explicit transfer-matrix elements and, consequently, equations~(\ref{eq:rs_general}) and~(\ref{eq:rc_from_rs}). The resulting values are the four ordered edges
\begin{equation}
 \kappa=2j+a,\qquad
 \kappa=2j+b,\qquad
 \kappa=2(j+1)-b,\qquad
 \kappa=2(j+1)-a,
 \label{eq:density_section_edges}
\end{equation}
with $a$ and $b$ defined in equation~(\ref{eq:edge_parameters}).  Thus every band edge is a state with a pure regular ($\sim|x-x_{sing}|^{1/2+\nu}$) or irregular ($\sim|x-x_{sing}|^{1/2-\nu}$) behavior at each singularity.  As shown in subsection~\ref{subsec:double_well_correspondence},
the same values constitute the union of the two scale-invariant spectra of the finite double-well problem.  The isolated fixed-point levels therefore become the endpoints of the Bloch bands once neighboring double wells are
coupled.

For the figures below, each Bloch state is normalized over one primitive
cell.

\begin{figure}[H]
 \centering
 \includegraphics[width=\textwidth]{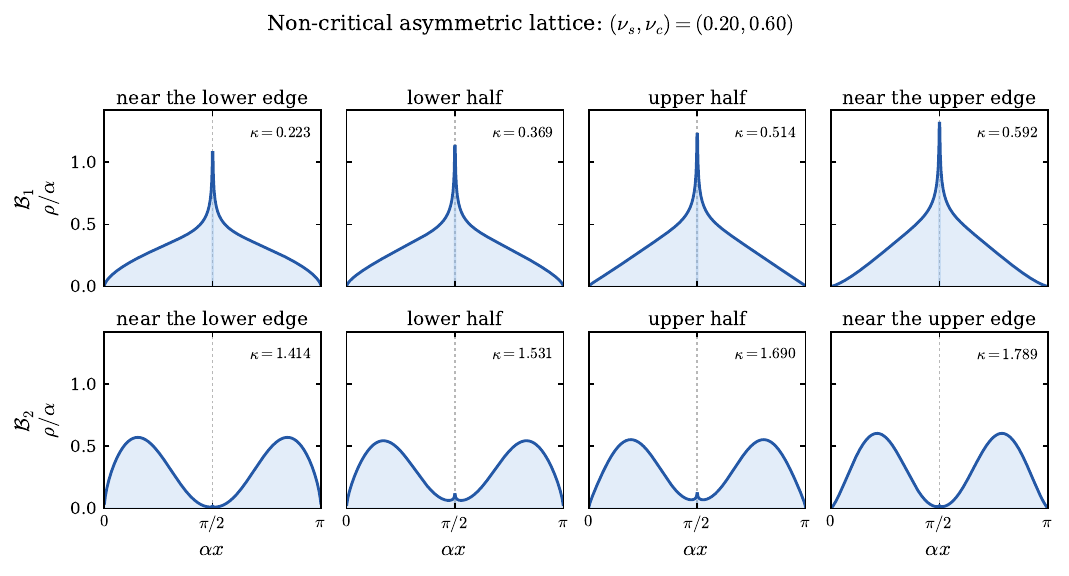}
 \caption{Normalized single-particle probability densities in one primitive
 cell of a non-critical asymmetric lattice with
 $(\nu_s,\nu_c)=(0.20,0.60)$. The upper and lower rows correspond to the first
 and second Bloch bands, respectively. From left to right, the states lie at
 $3\%$, $30\%$, $70\%$, and $97\%$ of the corresponding energy interval,
 measured from its lower edge. The dashed line at $\alpha x=\pi/2$ marks the
 $c$-type singularity, while the cell boundaries are $s$-type singularities.
 All densities are normalized and share a common vertical
 scale.}
 \label{fig:density_noncritical}
\end{figure}

\subsection{Lattice regimes}
\label{subsec:density_regimes}

\subsubsection{Non-critical asymmetric lattice}

When both $|\nu_s-\nu_c|$ and $|1-\nu_s-\nu_c|$ are appreciably different
from zero, equation~(\ref{eq:rs_general}) is the most economical expression
for the eigenvector. The first two bands contain the complete alternating
pattern of regular and irregular endpoint conditions; higher pairs repeat this
pattern while acquiring additional nodes inside the cell. We therefore
restrict the graphical analysis to the first two bands. Four states are shown
inside each band: one near the lower edge, two at intermediate energies below
and above the midpoint, and one near the upper edge. All four energies are
chosen strictly inside the open band, rather than at the isolated fixed-point
levels that form its limiting edges.
Figure~\ref{fig:density_noncritical} reveals a robust feature of the Bloch
states. As the energy is varied within a given band, the density changes
continuously, but its spatial organization does not change qualitatively.
The positions of the dominant maxima and minima, as well as the character of
the density near each singularity, remain the same; only the relative heights
and widths of the peaks evolve. The more substantial redistribution occurs
when the gap is crossed. For $(\nu_s,\nu_c)=(0.20,0.60)$, the first band is
strongly enhanced at the central $c$-type singularity, whereas the second
band develops two broad maxima in the intervals adjacent to the $s$-type
sites. Since $g_i=\nu_i^2-1/4$ increases monotonically with $\nu_i$, the
$s$ sites are the more attractive ones in this example. The second-band
density is consequently displaced toward the neighborhoods of the
smaller-$\nu$ singularities, while the first band is dominated by the
complementary irregular behavior at the larger-$\nu$ site.

\begin{figure}[H]
 \centering
 \includegraphics[width=\textwidth]
 {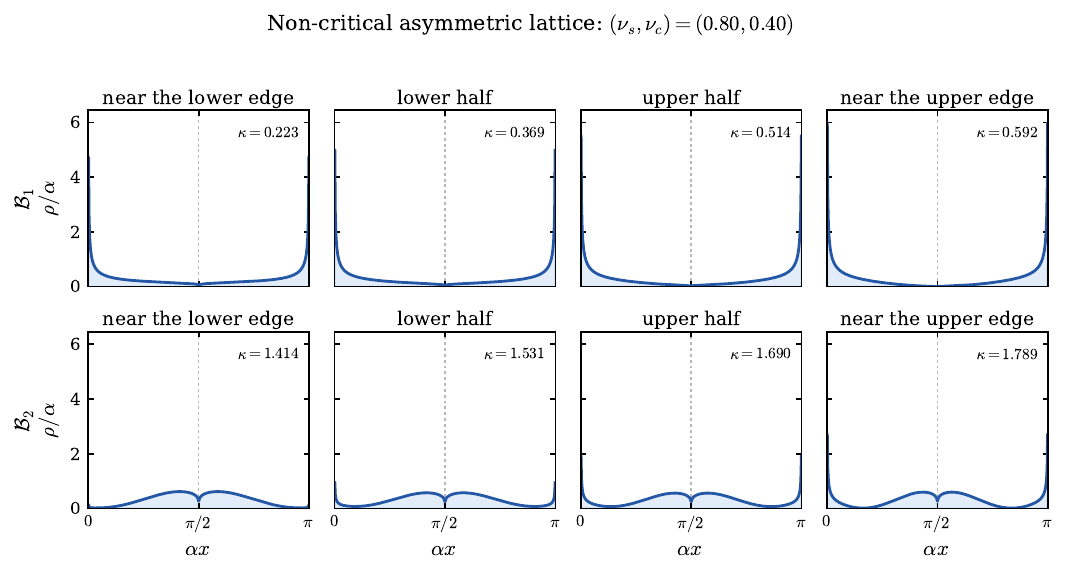}
 \caption{Normalized probability densities for the non-critical lattice
 $(\nu_s,\nu_c)=(0.80,0.40)$, which is isospectral to the lattice in
 figure~\ref{fig:density_noncritical}. Corresponding panels in the two
 figures have the same energy, $\kappa$, and Bloch momentum $Q$. The rows and
 columns follow the same convention as in figure~\ref{fig:density_noncritical}.
 Despite the identical band structure, the probability is redistributed
 between the $s$-type cell boundaries and the central $c$-type singularity.
 All panels share a common vertical scale within this figure.}
 \label{fig:density_noncritical_isospectral}
\end{figure}

This regime also provides the clearest illustration of the distinction
between spectral and wave-function symmetries. The discriminant is invariant
under the simultaneous transformation
\begin{equation}
 (\nu_s,\nu_c)\longrightarrow(1-\nu_s,1-\nu_c),
 \label{eq:isospectral_complement}
\end{equation}
but the local powers, coefficient ratios, and probability densities are not.
Thus $(\nu_s,\nu_c)=(0.20,0.60)$ and $(0.80,0.40)$ have exactly the same
band edges and dispersion relation. In particular, the values of $\kappa$
and $Q$ represented in corresponding panels of
figures~\ref{fig:density_noncritical} and
\ref{fig:density_noncritical_isospectral} are identical. Their spatial
profiles, however, are markedly different. Under the transformation, the
attractive member of the alternating pair changes from the $s$ site to the
$c$ site. Correspondingly, the dominant first-band peak moves from the
central singularity to the cell boundaries, while the broad second-band
weight is redirected toward the central, smaller-$\nu$ site. The comparison
therefore shows explicitly that a symmetry of the Bloch spectrum need not be
a symmetry of the Bloch eigenfunctions or of local observables derived from
them.

\subsubsection{Symmetric lattice}

For $\nu_s=\nu_c\equiv\nu$, the primitive period is
$\ell=\pi/(2\alpha)$, and the relevant one-step transfer matrix is
$\eta\mathcal C_{\mathrm{sym}}(\nu;\kappa)$.  Its two diagonal entries are equal, and the
coefficient ratio reduces to
\begin{equation}
 \alpha^{2\nu}\frac{B_0}{A_0}
 =\pm
 \left[
 -\frac{
 \Gamma(\nu)\Gamma(1+\nu)
 \Gamma\left(\frac12-\nu+\frac\kappa2\right)
 \Gamma\left(\frac12-\nu-\frac\kappa2\right)}
 {
 \Gamma(-\nu)\Gamma(1-\nu)
 \Gamma\left(\frac12+\nu+\frac\kappa2\right)
 \Gamma\left(\frac12+\nu-\frac\kappa2\right)}
 \right]^{1/2}.
 \label{eq:symmetric_B_over_A_density}
\end{equation}
Equivalently, the sign associated with a definite Bloch momentum is fixed by
\begin{equation}
 \alpha^{2\nu}\frac{B_0}{A_0}
 =-\frac{i\sin(Q\ell)}{C^{\mathrm{sym}}_{12}(\nu;\kappa)}.
 \label{eq:symmetric_B_over_A_phase}
\end{equation}
At the other end of the elementary interval,
\begin{equation}
 \alpha^{2\nu}\frac{B_1}{A_1}
 =-\alpha^{2\nu}\frac{B_0}{A_0}.
 \label{eq:symmetric_endpoint_ratios}
\end{equation}
Its zeros give $B_0=0$ and $\kappa=2n+1+2\nu$, whereas its poles give $A_0=0$ and the complementary fixed-point sequence.  The probability density has the shorter periodicity
$\varrho(x+\ell)=\varrho(x)$.  Accordingly, the first and second bands of the symmetric lattice should be understood as genuine bands in the unfolded Brillouin zone, rather than as the two folded branches obtained with the doubled asymmetric cell.

\begin{figure}[H]
 \centering
 \includegraphics[width=\textwidth]{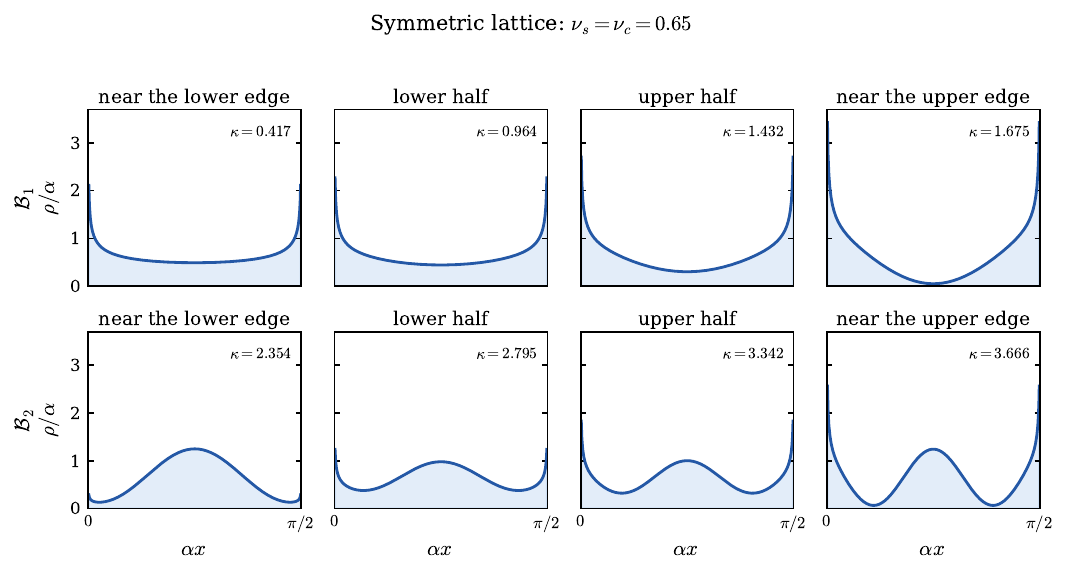}
 \caption{Normalized single-particle probability densities in one primitive
 cell of the symmetric lattice with $\nu_s=\nu_c=0.65$. The upper and lower
 rows show the first and second Bloch bands. From left to right, the states
 lie at $3\%$, $30\%$, $70\%$, and $97\%$ of the corresponding energy
 interval, measured from its lower edge. Here the primitive cell is
 $0<\alpha x<\pi/2$, with singularities at its two boundaries. All panels
 share a common vertical scale.}
 \label{fig:density_symmetric}
\end{figure}

The normalized densities in figure~\ref{fig:density_symmetric} make the
consequences of the shorter primitive cell explicit. In the first band the
weight remains concentrated near the singular boundaries, whereas in the
second band it is transferred predominantly toward the interior of the
cell. Once again, moving from one edge of a band to the other modifies the
sharpness and relative amplitude of the peaks without changing this basic
spatial pattern. The qualitative change is associated with the band index,
not with the position of the state inside a given band. Because the two
singularities are equivalent in the symmetric lattice, there is no preferred
site; the contrast appears instead as boundary-enhanced versus
interior-enhanced probability. This distinction follows from the alternating
regular and irregular fixed-point limits discussed above.

\subsubsection{Complementary couplings and the Dirac contact}

The second closing condition,
\begin{equation}
 \nu_s+\nu_c=1,
 \qquad \nu_s\neq\nu_c,
 \label{eq:complementary_density_line}
\end{equation}
does not reduce the primitive period.  In this case the transfer matrix again
has equal diagonal entries, even though the lattice remains asymmetric.  If
$\nu_s=\nu$ and $\nu_c=1-\nu$, equation~(\ref{eq:rs_general}) simplifies to
\begin{equation}
 \left(\alpha^{2\nu}\frac{B_0}{A_0}\right)^2
 =-\frac{4\nu^2}{\kappa^2}
 \left[\frac{\Gamma(\nu)}{\Gamma(1-\nu)}\right]^2
 \frac{\mathcal G_{2-2\nu}(\kappa)}
      {\mathcal G_{2\nu}(\kappa)},
 \label{eq:complementary_B_over_A}
\end{equation}
where $\mathcal G_z$ is defined in
appendix~\ref{app:transfer_eigenvectors}.  As in the symmetric case, the
choice between the two square roots is fixed by the sign of $Q$.

At the massless contacts, $\kappa=N=2j+2$ and $Q=0$, the
complete-period matrix becomes
\begin{equation}
 \mathcal T(N)=\mathbb I.
 \label{eq:T_identity_dirac_contact}
\end{equation}
If the contact point is considered in isolation, every coefficient vector
is therefore a Bloch eigenvector with unit eigenvalue, and the ratio
$B_0/A_0$ appears to be undetermined. The surrounding band structure,
however, selects a natural pair of states. Writing $\nu_s=\nu$ and
$\nu_c=1-\nu$, the two finite limiting values are
\begin{equation}
 \lim_{Q\rightarrow0}
 \alpha^{2\nu}\frac{B_0}{A_0}
 =
 \pm i\,\mathcal R_N,
 \qquad
 \mathcal R_N=
 \frac{2\nu}{N}
 \frac{\Gamma(\nu)}{\Gamma(1-\nu)}
 \frac{
 \Gamma\left(\frac{N}{2}+1-\nu\right)}
 {\Gamma\left(\frac{N}{2}+\nu\right)}.
 \label{eq:Dirac_contact_eigenvectors}
\end{equation}
More precisely, as $Q\rightarrow0^+$, the lower and upper bands approach $+i\mathcal R_N$ and $-i\mathcal R_N$, respectively, whereas for $Q\rightarrow0^-$ these assignments are reversed. These results show that the coefficient ratio is continuous when the
contact is crossed along either of the two Dirac trajectories. Following the trajectory from the lower band at $Q<0$ to the upper band at $Q>0$ gives the same limiting value, $-i\mathcal R_N$, on both sides of the contact. Following the opposite trajectory, from the lower band at $Q>0$ to the upper band at $Q<0$, gives instead $+i\mathcal R_N$ on both sides. Thus, although the upper- and lower-band labels are exchanged at $Q=0$, each trajectory selects a definite and continuous value of $B_0/A_0$.

\begin{figure}[H]
 \centering
 \includegraphics[width=\textwidth]{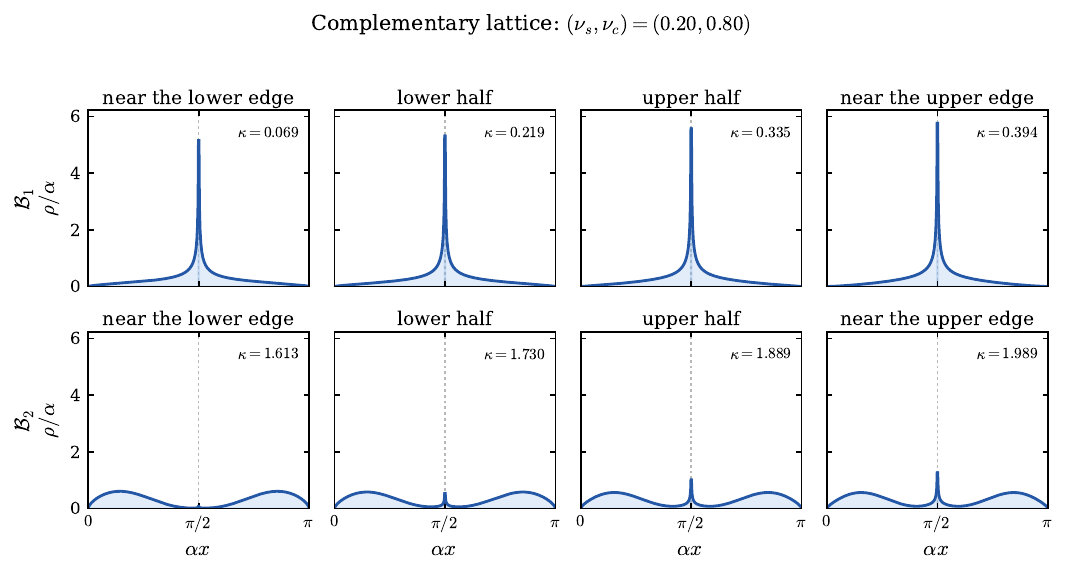}
 \caption{Normalized single-particle probability densities in one primitive
 cell for complementary couplings $(\nu_s,\nu_c)=(0.20,0.80)$, for which
 $\nu_s+\nu_c=1$. The upper and lower rows correspond to the first and second
 Bloch bands. From left to right, the states lie at $3\%$, $30\%$, $70\%$,
 and $97\%$ of the corresponding energy interval, measured from its lower
 edge. The dashed line at $\alpha x=\pi/2$ marks the $c$-type singularity.
 No state is evaluated exactly at $Q=0$; the final state of the second band
 lies close to the massless contact. A common vertical scale is used in all
 panels.}
 \label{fig:density_complementary}
\end{figure}
Figure~\ref{fig:density_complementary} shows the corresponding probability
redistribution in the first two bands. The plotted states remain strictly
inside the bands; in particular, the state closest to the massless contact
is taken at a small nonzero value of $Q$. This selects one of the two
continuous Dirac trajectories and avoids the basis ambiguity of the exactly
degenerate point, while displaying its limiting spatial profile. The same
bandwise stability found in the non-critical and symmetric lattices persists
here. Throughout the first band the density is dominated by the central
$c$-type singularity, which has the larger value of $\nu$. In the second
band the broad part of the probability distribution is shifted toward the
two $s$-type sites, for which $\nu_s$ is smaller and the interaction is more
attractive. A progressively sharper central contribution becomes visible as
the Dirac contact is approached, but it does not alter the overall continuity
of the density along the selected branch.

Taken together, figures~\ref{fig:density_noncritical}--
\ref{fig:density_complementary} separate two effects that are not visible in
the dispersion relation alone. Within each band, changing $Q$ produces a
smooth quantitative deformation of an essentially fixed probability
profile. Crossing a gap and entering the next band changes the asymptotic
sector and produces a qualitatively different spatial distribution. In an
asymmetric cell this usually appears as an alternation between a state
controlled by the larger-$\nu$ singularity and a state whose broad weight is
drawn toward the smaller-$\nu$, more attractive (or less repulsive) site.
The isospectral pair further demonstrates that even an identical set of Bloch
bands does not determine this spatial information.

%%%%%%%%%%%%%%%%%%%%%%%%%%%%%%%%%%%%%%%%%%%
\section{Conclusions}
\label{sec:conclusions}

We have constructed and analyzed an exactly solvable one-dimensional lattice
whose elementary cells are separated by inverse-square singularities. The
central point is that such singularities do not necessarily represent
impenetrable walls. In the weak-to-medium coupling regime
$0<\nu_s,\nu_c<1$, both local solutions are square integrable, and square integrability alone does not determine the domain of the Hamiltonian. The conventional choice that retains only the less singular solution would
disconnect the cells and reduce the periodic system to a collection of
independent wells. Here, instead, a current-preserving matching condition was
used to connect the two admissible behaviors across every singular point.
This prescription defines a particular self-adjoint extension of the
Hamiltonian on the full line. Bloch's condition can then be imposed on the
resulting global problem, rather than independently within isolated cells.

The local asymptotic bases were normalized so that the boundary form has a
fixed canonical normalization. This made it possible to encode the matching
and the exact hypergeometric connection formulas in a unimodular transfer
matrix and to determine the Bloch spectrum in closed form for both the
asymmetric and symmetric lattices. The band edges were shown to coincide with
the union of the two scale-invariant spectra of the associated
finite double-well problem. Thus the isolated fixed-point levels become the
endpoints of finite Bloch bands once transmission through the singularities
is allowed. This provides a direct connection between the self-adjoint
boundary data of the singular well and the band structure of its periodic
extension.

The exact spectrum also revealed two independent mechanisms for closing a
gap. When $|\nu_s-\nu_c|\rightarrow0$, the two singular sublattices become
equivalent and the primitive period is reduced by one half. The corresponding
Brillouin zone must therefore be unfolded, and the apparent crossing in the
doubled-cell description is a consequence of band folding. By contrast, on
the complementary-coupling line $\nu_s+\nu_c=1$ with
$\nu_s\neq\nu_c$, the primitive period remains unchanged and a different
family of gaps closes. The resulting linear contact is therefore a genuine
massless Dirac-like cone in the original Brillouin zone. Away from these
closing conditions, ordinary quadratic band edges define effective electron
and hole masses. Close to a small but finite gap, the exact dispersion also
exhibits a crossover between the conventional quadratic regime and a massive
Dirac-like regime, with the gap playing the role of an effective Dirac mass.

The Bloch eigenvectors contain information that is absent from the discriminant. We found that the normalized probability density changes smoothly but not qualitatively as the energy is varied within a given band. The main spatial redistribution occurs when a gap is crossed and the next band is entered. In an asymmetric cell,
this produces an alternation between states controlled by the larger-$\nu$
singularity and states whose broad probability weight is drawn toward the
smaller-$\nu$, more attractive or less repulsive site. Moreover, the
transformation $(\nu_s,\nu_c)\mapsto(1-\nu_s,1-\nu_c)$ leaves the complete
band structure invariant while changing the local powers and the probability
density. Hence spectral equivalence does not imply equivalence of the Bloch
eigenfunctions or of local observables.

From a condensed-matter perspective, the model provides a one-dimensional setting in which two Bloch bands intersect linearly. Near the contact, the two counterpropagating branches display the characteristic dispersion of
massless Dirac modes, even though the underlying model remains a single-component nonrelativistic Schr\"odinger system. Tuning the singular couplings opens, reduces, or closes the corresponding gap, thereby producing a continuous crossover between massive and massless Dirac-like band dispersions.
%From a condensed-matter perspective, the model provides a one-dimensional setting in which two bands intersect linearly and the low-energy excitations are described by counterpropagating massless Dirac modes, an importantkinematic feature of low-dimensional systems~\cite{GarreauZehnle2017,ZeunerEtAl2012}. Tuning the singular couplings opens, reduces, or closes the associated Dirac-like gap, thereby emulating the transition between massive and massless quasiparticles. 

At the same time, the exact solution retains control of the complete spectrum and wave functions, including the quadratic corrections away from the crossing. Although the model is most naturally regarded as an idealized lattice rather than as a literal electronic crystal, its one-dimensional character does not make the construction physically empty. Effective Dirac dynamics and tunable band crossings can be engineered in quasi-one-dimensional settings, including ultracold atoms in driven optical lattices and photonic waveguide arrays~\cite{GarreauZehnle2017,ZeunerEtAl2012}. In such platforms the inverse-square poles may be viewed as ideal
limits of sharply structured barriers or trapping regions. The present
solution can therefore serve both as a benchmark for regularized models and
as a quantum-simulation template in which transmission, gap opening,
effective masses, and the spatial redistribution of Bloch states can be
investigated independently. A direct connection with electronic conductivity
would require additional ingredients, including filling, scattering, and
interactions, which lie beyond the single-particle analysis developed here.

Finally, the transmitting condition adopted in this work is only one member
of the family of self-adjoint extensions admitted by the inverse-square
singularities. Other choices may relate the local asymptotic coefficients in
different ways and may introduce an additional length scale. Consequently,
even the same trigonometric P\"oschl--Teller differential expression can
support inequivalent lattice realizations with distinct spectral and spatial
properties. Moreover, the extension data need not be identical at the two
singular sublattices or at every cell. A periodic modulation of the matching
conditions over several cells would define a supercell and hence a
superlattice, whereas a localized change could emulate an isolated defect or
impurity. The self-adjoint extension parameters can thus be regarded as
physical design variables that allow the same solvable potential to represent
different lattice environments.

This flexibility is not restricted to the trigonometric P\"oschl--Teller
potential, since the local matching mechanism depends only on the leading
inverse-square behavior near each singularity. Other exactly solvable
singular systems can be periodically extended using the same structure,
while their model-dependent connection matrices encode the propagation
between neighboring singularities. A natural next example is the
trigonometric Eckart potential, which combines a
$1/\sin^2(\alpha x)$ term with a $\cot(\alpha x)$ contribution. The latter is
subleading near each inverse-square pole and changes the propagation through
the cell without altering the local classification responsible for the
matching. Applying the present construction to that model should produce an
exactly tractable lattice with intrinsically asymmetric cells. Transport and
many-particle observables constitute further directions in which this family
of solvable singular lattices may be explored.

%%%%%%%%%%%%%%%%%%%%%%%%%%%%
%\section*{Acknowledgments}
%The author thanks G.~M.~Sotkov for his critical reading of the manuscript.

\section*{Funding}
This research received no specific funding.

\section*{Author contributions}
The sole author conceived the study, performed the analytical
calculations, prepared the figures, and wrote the manuscript.

\section*{Data availability}
All results supporting the conclusions of this study,
including the analytical expressions used to generate
the figures, are contained within the article.

\section*{Supplementary data}
No supplementary material accompanies this article.
%%%%%%%%%%%%
\clearpage
\appendix
\section{Connection matrices and Bloch eigenvectors}
\label{app:transfer_eigenvectors}

This appendix collects the explicit connection and transfer matrices used in
the main text.  We adopt throughout the matching convention
\begin{equation}
 A_{>}(k)=-A_{<}(k),
 \qquad
 B_{>}(k)=B_{<}(k),
 \qquad
 \eta=\operatorname{diag}(-1,1),
 \label{eq:app_matching_convention}
\end{equation}
so that the $A$ solution is extended oddly across each singularity, whereas
the $B$ solution is extended evenly.  It is useful to define
\begin{equation}
 \mathcal G_z(\kappa)\equiv
 \Gamma\!\left(\frac{z+\kappa}{2}\right)
 \Gamma\!\left(\frac{z-\kappa}{2}\right),
 \qquad
 \kappa=\frac{k}{\alpha}.
 \label{eq:app_gamma_product}
\end{equation}

\subsection{Connection matrices}

The connection matrix from an $s$ singularity to the neighboring $c$
singularity is defined by
\begin{equation}
 \begin{pmatrix}
  \alpha^{-\nu_c}A_1(k)\\[1mm]
  \alpha^{\nu_c}B_1(k)
 \end{pmatrix}
 =\mathcal C_{sc}(\kappa)
 \begin{pmatrix}
  \alpha^{-\nu_s}A_0(k)\\[1mm]
  \alpha^{\nu_s}B_0(k)
 \end{pmatrix},
 \qquad
 \mathcal C_{sc}(\kappa)=
 \begin{pmatrix}
  C^{sc}_{11}&C^{sc}_{12}\\
  C^{sc}_{21}&C^{sc}_{22}
 \end{pmatrix}.
 \label{eq:app_Csc_definition}
\end{equation}
Its entries are
\begin{align}
 C^{sc}_{11}={}&
 \sqrt{\frac{\nu_c}{\nu_s}}
 \frac{\Gamma(-\nu_c)\Gamma(1+\nu_s)}
      {\mathcal G_{1+\nu_s-\nu_c}(\kappa)},
 &
 C^{sc}_{12}={}&
 \sqrt{\frac{\nu_c}{\nu_s}}
 \frac{\Gamma(-\nu_c)\Gamma(1-\nu_s)}
      {\mathcal G_{1-\nu_s-\nu_c}(\kappa)},
 \nonumber\\[2mm]
 C^{sc}_{21}={}&
 \sqrt{\frac{\nu_c}{\nu_s}}
 \frac{\Gamma(\nu_c)\Gamma(1+\nu_s)}
      {\mathcal G_{1+\nu_s+\nu_c}(\kappa)},
 &
 C^{sc}_{22}={}&
 \sqrt{\frac{\nu_c}{\nu_s}}
 \frac{\Gamma(\nu_c)\Gamma(1-\nu_s)}
      {\mathcal G_{1-\nu_s+\nu_c}(\kappa)}.
 \label{eq:app_Csc_entries}
\end{align}

For the second half of the primitive cell, the two singularities are
interchanged:
\begin{equation}
 \mathcal C_{cs}(\kappa)=
 \left.\mathcal C_{sc}(\kappa)\right|_{\nu_s\leftrightarrow\nu_c}
 =
 \begin{pmatrix}
  C^{cs}_{11}&C^{cs}_{12}\\
  C^{cs}_{21}&C^{cs}_{22}
 \end{pmatrix},
 \label{eq:app_Ccs_definition}
\end{equation}
where
\begin{align}
 C^{cs}_{11}={}&
 \sqrt{\frac{\nu_s}{\nu_c}}
 \frac{\Gamma(-\nu_s)\Gamma(1+\nu_c)}
      {\mathcal G_{1+\nu_c-\nu_s}(\kappa)},
 &
 C^{cs}_{12}={}&
 \sqrt{\frac{\nu_s}{\nu_c}}
 \frac{\Gamma(-\nu_s)\Gamma(1-\nu_c)}
      {\mathcal G_{1-\nu_s-\nu_c}(\kappa)},
 \nonumber\\[2mm]
 C^{cs}_{21}={}&
 \sqrt{\frac{\nu_s}{\nu_c}}
 \frac{\Gamma(\nu_s)\Gamma(1+\nu_c)}
      {\mathcal G_{1+\nu_s+\nu_c}(\kappa)},
 &
 C^{cs}_{22}={}&
 \sqrt{\frac{\nu_s}{\nu_c}}
 \frac{\Gamma(\nu_s)\Gamma(1-\nu_c)}
      {\mathcal G_{1-\nu_c+\nu_s}(\kappa)}.
 \label{eq:app_Ccs_entries}
\end{align}
The Gamma-function recurrence and reflection identities give
\begin{equation}
 \det\mathcal C_{sc}=\det\mathcal C_{cs}=-1.
 \label{eq:app_C_determinants}
\end{equation}

\subsection{Asymmetric lattice}

For $\nu_s\neq\nu_c$, the primitive period is $L=\pi/\alpha$ and the
transfer matrix is
\begin{equation}
 \mathcal T(\kappa)
 =\eta\mathcal C_{cs}(\kappa)\eta\mathcal C_{sc}(\kappa)
 =
 \begin{pmatrix}
  \mathcal T_{11}&\mathcal T_{12}\\
  \mathcal T_{21}&\mathcal T_{22}
 \end{pmatrix}.
 \label{eq:app_T_definition}
\end{equation}
Writing $\mathcal C_{sc}=\left(\begin{smallmatrix}a&b\\c&d\end{smallmatrix}\right)$
and $\mathcal C_{cs}=\left(\begin{smallmatrix}a'&b'\\c'&d'\end{smallmatrix}\right)$,
one has
\begin{align}
 \mathcal T_{11}&=a'a-b'c,
 &
 \mathcal T_{12}&=a'b-b'd,
 \nonumber\\
 \mathcal T_{21}&=-c'a+d'c,
 &
 \mathcal T_{22}&=-c'b+d'd.
 \label{eq:app_T_entries_algebraic}
\end{align}
In terms of $\nu_s$ and $\nu_c$, these elements reduce to
\begin{align}
 \mathcal T_{11}=\mathcal T_{22}={}&
 \frac{\cos(\pi\kappa)+\cos(\pi\nu_s)\cos(\pi\nu_c)}
      {\sin(\pi\nu_s)\sin(\pi\nu_c)},
 \label{eq:app_T11_T22}\\[2mm]
 \mathcal T_{12}={}&
 \frac{2\nu_c\,
  \Gamma(-\nu_s)\Gamma(1-\nu_s)
  \Gamma(\nu_c)\Gamma(-\nu_c)}
 {\mathcal G_{1+\nu_c-\nu_s}(\kappa)
  \mathcal G_{1-\nu_s-\nu_c}(\kappa)},
 \label{eq:app_T12}\\[2mm]
 \mathcal T_{21}={}&-
 \frac{2\nu_c\,
  \Gamma(\nu_s)\Gamma(1+\nu_s)
  \Gamma(\nu_c)\Gamma(-\nu_c)}
 {\mathcal G_{1+\nu_s+\nu_c}(\kappa)
  \mathcal G_{1+\nu_s-\nu_c}(\kappa)}.
 \label{eq:app_T21}
\end{align}
Consequently,
\begin{equation}
 \det\mathcal T=1,
 \qquad
 \mathcal T_{12}\mathcal T_{21}=\mathcal T_{11}^{2}-1.
 \label{eq:app_T_properties}
\end{equation}

Defining the dimensionless coefficient ratio
\begin{equation}
 r_s=\alpha^{2\nu_s}\frac{B_0}{A_0},
 \label{eq:app_rs_definition}
\end{equation}
and solving the Bloch eigenvalue equation directly for $r_s$ gives
\begin{equation}
 \mathcal T_{12}r_s^2+
 (\mathcal T_{11}-\mathcal T_{22})r_s-
 \mathcal T_{21}=0.
 \label{eq:app_ratio_quadratic}
\end{equation}
The two solutions, associated with the Bloch phases $e^{\pm iQL}$, are
\begin{equation}
 r_s^{(\pm)}=
 \frac{e^{\pm iQL}-\mathcal T_{11}}{\mathcal T_{12}}
 =\frac{\mathcal T_{21}}
 {e^{\pm iQL}-\mathcal T_{22}}.
 \label{eq:app_rs_bloch_phase}
\end{equation}

\subsection{Symmetric lattice}

When $\nu_s=\nu_c\equiv\nu$, the primitive period is
$\ell=\pi/(2\alpha)$.  The connection matrix becomes
\begin{equation}
 \mathcal C_{\mathrm{sym}}(\nu;\kappa)=
 \begin{pmatrix}
 \displaystyle
 -\frac{\cos(\pi\kappa/2)}{\sin(\pi\nu)}
 &
 \displaystyle
 \frac{\Gamma(-\nu)\Gamma(1-\nu)}
      {\mathcal G_{1-2\nu}(\kappa)}
 \\[5mm]
 \displaystyle
 \frac{\Gamma(\nu)\Gamma(1+\nu)}
      {\mathcal G_{1+2\nu}(\kappa)}
 &
 \displaystyle
 \frac{\cos(\pi\kappa/2)}{\sin(\pi\nu)}
 \end{pmatrix}.
 \label{eq:app_symmetric_C}
\end{equation}
With the convention in equation~(\ref{eq:app_matching_convention}), the
one-step transfer matrix is
\begin{equation}
 \mathcal M_{\mathrm{sym}}(\nu;\kappa)
 =\eta\mathcal C_{\mathrm{sym}}(\nu;\kappa)
 =
 \begin{pmatrix}
 \displaystyle
 \frac{\cos(\pi\kappa/2)}{\sin(\pi\nu)}
 &
 \displaystyle
 -\frac{\Gamma(-\nu)\Gamma(1-\nu)}
       {\mathcal G_{1-2\nu}(\kappa)}
 \\[5mm]
 \displaystyle
 \frac{\Gamma(\nu)\Gamma(1+\nu)}
       {\mathcal G_{1+2\nu}(\kappa)}
 &
 \displaystyle
 \frac{\cos(\pi\kappa/2)}{\sin(\pi\nu)}
 \end{pmatrix}.
 \label{eq:app_symmetric_M}
\end{equation}
Its trace gives
\begin{equation}
 \cos(Q\ell)=
 \frac{\cos(\pi\kappa/2)}{\sin(\pi\nu)},
\end{equation}
and its equal diagonal entries imply
\begin{equation}
 \left(\alpha^{2\nu}\frac{B_0}{A_0}\right)^2
 =-
 \frac{\Gamma(\nu)\Gamma(1+\nu)}
      {\Gamma(-\nu)\Gamma(1-\nu)}
 \frac{\mathcal G_{1-2\nu}(\kappa)}
      {\mathcal G_{1+2\nu}(\kappa)}.
 \label{eq:app_symmetric_ratio}
\end{equation}

\subsection{Complementary-coupling line}

Let $\nu_s=\nu$ and $\nu_c=1-\nu$.  The entries of
$\mathcal C_{sc}=\left(\begin{smallmatrix}a&b\\c&d\end{smallmatrix}\right)$
simplify to
\begin{align}
 a={}&-\sqrt{\frac{\nu}{1-\nu}}
 \frac{\Gamma^2(\nu)}{\mathcal G_{2\nu}(\kappa)},
 &
 d={}&\sqrt{\frac{1-\nu}{\nu}}
 \frac{\Gamma^2(1-\nu)}
      {\mathcal G_{2-2\nu}(\kappa)},
 \nonumber\\[2mm]
 b={}&
 \frac{\kappa\sin(\pi\kappa/2)}
 {2\sin(\pi\nu)\sqrt{\nu(1-\nu)}},
 &
 c={}&
 \frac{2\sqrt{\nu(1-\nu)}\sin(\pi\kappa/2)}
 {\kappa\sin(\pi\nu)}.
 \label{eq:app_complementary_C_entries}
\end{align}
The reversed connection matrix is
\begin{equation}
 \mathcal C_{cs}=
 \begin{pmatrix}
  -d&b\\ c&-a
 \end{pmatrix},
\end{equation}
and therefore
\begin{equation}
 \mathcal T_{11}=\mathcal T_{22}=-(ad+bc),
 \qquad
 \mathcal T_{12}=-2bd,
 \qquad
 \mathcal T_{21}=-2ac.
 \label{eq:app_complementary_T_entries}
\end{equation}
It follows that
\begin{equation}
 \left(\alpha^{2\nu}\frac{B_0}{A_0}\right)^2
 =\frac{\mathcal T_{21}}{\mathcal T_{12}}
 =-\frac{4\nu^2}{\kappa^2}
 \left[\frac{\Gamma(\nu)}{\Gamma(1-\nu)}\right]^2
 \frac{\mathcal G_{2-2\nu}(\kappa)}
      {\mathcal G_{2\nu}(\kappa)}.
 \label{eq:app_complementary_ratio}
\end{equation}
At $\kappa=2j+2$, both off-diagonal entries of $\mathcal T$ vanish and the
diagonal entries tend to unity.  Thus $\mathcal T=\mathbb I$ at each massless
contact.  Equation~(\ref{eq:app_complementary_ratio}) must then be understood
as the ratio selected by approaching the contact along either of the two
bands; at the degenerate point itself, arbitrary linear combinations of the
two independent periodic solutions are also allowed.

\bibliographystyle{unsrt}
\bibliography{PT_lattice_references}

\end{document}